\documentclass[a4paper,12pt]{article}
\usepackage{a4wide}
\usepackage[bookmarks=false,colorlinks]{hyperref}
\usepackage{graphicx,graphics,epsfig,color}
\usepackage{amsmath,amsfonts,amssymb,mathtools,bm}
\usepackage{rotating}
\usepackage{tabularx}
\usepackage{multicol,multirow}
\usepackage{latexsym}
\usepackage{relsize}
\usepackage{subfigure}
\usepackage{float}
\usepackage{slashed}
\usepackage{lineno}
\usepackage{cite}
\usepackage{appendix}
\usepackage{url}

\date{\today}
 
\normalsize

\hypersetup{urlcolor=blue, citecolor=blue}

\begin{document}

%\renewcommand{\thefootnote}{\fnsymbol{footnote}}

%\leftline\today
\rightline{LPT-ORSAY-11-100}
\rightline{LAL-11-307}
\vspace{1cm} 
{\Large
\begin{center}
	{\bf The strong decays of $K_1$ resonances}
\end{center}
}
\vspace{0.3cm}

\begin{center}
	A. Tayduganov$^{a,b}$, E. Kou$^a$, A. Le Yaouanc$^b$\\
	\vspace{0.3cm}\small
	\emph{$^a$ Laboratoire de l'Acc\'el\'erateur Lin\'eaire, Univ. Paris-Sud 11, CNRS/IN2P3 (UMR 8607)} \\
	\emph{91405 Orsay, France}

	\emph{$^b$ Laboratoire de Physique Th\'eorique, CNRS/Univ. Paris-Sud 11 (UMR 8627)}\\
	\emph{91405 Orsay, France}
\end{center}

\vskip1cm

\vskip1.3cm

\begin{center}
	\small{\bf Abstract}\\[3mm]
\end{center}

We investigate the $K_1\to K\pi\pi$ strong interaction decays. Using the $^3P_0$ quark-pair-creation model to derive the basic parametrization, we discuss in detail how to obtain the various partial wave amplitudes into the possible quasi-two-body decay channels as well as their relative phases from the currently available experimental data. We obtain the $K_1$ mixing angle to be $\theta_{K_1}\simeq 60^{\circ}$,  in agreement with previous works. Our study  can be applied to extract the information needed for the photon polarization determination of the radiative $B\to K_1\gamma$ decay.

\vskip0.3cm

%\newpage

\cleardoublepage
\tableofcontents
\cleardoublepage

\section{Introduction}

1 \underline{\it  Motivation for revisiting $K_1$-meson strong decays} \\
It has been proposed a method to measure the polarization of the photon in weak radiative decays of the $B$-meson by exploiting the decay $B\to K\pi\pi\gamma$, with the system $K\pi\pi$ resonating into a $K_1$ state~\cite{Gronau:2002rz,Gronau:2001ng}. We have recently extended this work~\cite{Kou:2010kn} and shown that exploiting the full Dalitz plot for the $K\pi\pi$ system could increase the sensitivity to the polarization determination  inspired by the  DDLR method~\cite{Davier:1992nw}. For this purpose, it is important to have a good understanding of the strong $K_1$-decays. Indeed, it turned out that the $B\to K_1(1270)\gamma$ channel, not considered in the original works~\cite{Gronau:2002rz,Gronau:2001ng}, dominates over $K_1(1400)$~\cite{Yang:2004as} while the pattern of the partial wave is especially complex for $K_1(1270)$.  

In the present paper, we give a detailed account and full discussion of the $K_1$ hadronic decays. 
The amplitude of the process $K_1\to K\pi\pi$ can be described by the basic quantity $\vec{\mathcal{J}}$:

\begin{equation}
	\mathcal{M}(K_1\to K\pi\pi)\propto\vec\varepsilon\cdot\vec{\mathcal{J}}
	\label{eq:J-function}
\end{equation}
where  $\vec\varepsilon$ is the polarization vector of the $K_1$ in the $K_1$ rest frame. 
The general framework for calculating $\vec{\mathcal{J}}$ is the \underline{\it quasi-two-body approximation}: the process $K_1\to K\pi\pi$ is decomposed into two steps: 1) the decay of $K_1\to$ vector isobar~($V$) + pseudoscalar~($P$); 2) the decay of the vector isobar ($K^{*}$ or $\rho$) into 2 pseudoscalars.
Then, $\mathcal{J}$ is a sum of terms which are products of couplings and one isobar denominator~\footnote{In the full expression for the weak process, $\vec\varepsilon\cdot\vec{\mathcal{J}}$ has still to be multiplied by a production amplitude and the corresponding Breit-Wigner denominator for the $K_1$.}. The explicit expressions for $\mathcal{J}$ have been given in \cite{Kou:2010kn}. The decay properties of the intermediate isobars are well known. Here, we are then interested in evaluating the couplings describing the first step of the decay, $K_1\to VP$, and the relative signs or phases between the various channels. 
Concerning our motivation,  it has appeared that the determination of the polarization parameter, called $\lambda_{\gamma}$, of $B\to K_1\gamma$ depends essentially on the expression  $Im[\vec{n}\cdot(\vec{\mathcal{J}}\times\vec{\mathcal{J}}^*)]/|\vec{\mathcal{J}}|^2$\cite{Gronau:2002rz,Gronau:2001ng,Kou:2010kn}. This expression vanishes unless complex phases are present in $\mathcal{J}$. These are mainly provided by the Breit-Wigner (BW) denominators of intermediate resonances (the so-called ``isobars''), and possibly by complex phases of the couplings. It is found that such a quantity is very sensitive to the relative signs of the various channels in the strong decay, whence it is important to determine the signs or, possibly, complex phases of the couplings, which anyway can be observed in the various experiments provided one measures a sufficient number of angular distributions. One requires also specifically good knowledge of 1) $D$-waves; 2) off-shell extrapolation.
\vspace*{0.4cm}

\noindent
2 \underline{\it  Status of experimental study of $K_1$-mesons}\\
 In principle, all the necessary hadronic parameters (i.e. $K_1$ masses and partial decay widths, form factors and relative phases) can be determined from fits to the experimental data. However, at present moment we are far from being able to perform this with good accuracy, the experiments suffer from many drawbacks.
The main reason is that the case under study cumulates many difficulties and complications, which have been underestimated in theoretical discussions: 
\begin{enumerate}
	\item several possible partial waves ($S,D$) for the same channel
	\item three-body decay through multiple interfering channels
	\item broad parent $K_1$ resonances
	\item effect of the large widths of isobars $K^*$ and $\rho$ 
	\item effect of a threshold ($K\rho$) close to the $K_1(1270)$ resonance
	\item overlapping and mixing of two close states, $K_1(1270)$ and $K_1(1400)$
\end{enumerate}

Perhaps not surprisingly for this particularly complicated case, one observes a rather confusing situation in experiments (e.g. contrary statements on total widths, on the $\kappa\pi$ decay channel, etc...). This has been causing misunderstandings on these  important observables. Then, one cannot simply use for example, the PDG entries, as is done usually. One should return to the original papers to understand what is actually measured, and this is not always easy. And also, we noticed some definite weak points. Up to now the most complete and accurate experimental analysis is the one by Daum {\it et al.}. It relies unhappily on the problematic phase space treatment of Nauenberg and Pais for decays to isobars. One notes also the absence of non resonant background in the $K$-matrix, and the presence of unexplained ``offset'' phases. All this is explained in details in section~\ref{sec:problems}. Finally, there is also lack of important informations like conventions of coupling signs or incomplete report of the parameters of the fit. Other experiments, which have been mentioned above, give precious complementary information but they are not able to solve all the problems, all the more since they are less accurate, and include less physical features in their fits  (e.g. neglect of $D$-waves). Then, a sizable part of our work has consisted in an extensive discussion of the experimental analyses (mainly the one of Daum {\it et al.} and the one of Belle).

\vspace*{0.4cm}

\noindent
3 \underline{\it  The theoretical treatments of $K_1$-mesons} \\
The theoretical model has a first aim to give a physical understanding of the observed decay properties, which are far from trivial. In addition, it may serve to complement the experimental knowledge, where there are some lacks or weaknesses, and to help for future experimental analyses. Of course, there is no fundamental theoretical treatment of such processes. We have only \underline{\it phenomenological approaches} at our disposal, mainly the one provided by the quark models. Approximate as it is, the quark model  can be very precious to check the consistency of the present data and to orient the future studies of $K_1$-decays. However, we have to keep in mind that it suffers from inherent, sizable and unknown uncertainties, which could limit the accuracy of the $\lambda_\gamma$ determination {as mentioned earlier}. To our knowledge, the best phenomenological model in the present case of $K_1$-decays is the \underline{\it family of quark models}, which are able to master the large set of hadronic states and their decays with a limited number of adjusted parameters~\footnote{See~\cite{Roca:2003uk} for another approach based on phenomenological Lagrangian.}. 

Admittedly, quark models are many, but one must distinguish between the potential models, and the decay models. The diversity  is especially the one of potential models, which intend to describe the spectroscopy of states. As concerns decays, there are not so many basic models. In fact, there are elementary emission models and quark-pair-creation models, both concerning two-body decays. This is why the quasi-two-body decay assumption is a natural step in the theoretical treatment of the three-body decay. The quark-pair-creation models have the advantage of unifying the whole of two-body and quasi-two-body decays. Among them, the  $^3P_0$ model (see section \ref{sec:QPCM} and references therein) is particularly favoured as being the easiest to handle, and then the more extensively tested, with a striking overall success over hundreds of decays. We use the $^3P_0$ model with the important additional input of a damping factor, to account for off-shellness. Of course, as said before, quark models are inherently approximate. As to the proper decay model, the main problem is that it is essentially non-relativistic, which is of course in principle very far from the real situation. It is known from quite a long time that quite surprisingly, non-relativistic decay or emission models may work well, but their accuracy cannot be estimated a priori, it has always to be judged a posteriori. Decay models must be necessarily combined with potential models giving the wave functions that must be folded into their general structure. In view of the rather naive status of the  $^3P_0$ model, we do not find it appropriate to use a sophisticated set of wave functions, but rather a simple-minded one, as explained in section \ref{sec:QPCM}. We must underline however that the oscillator radii that are used are not at all free parameters: they have to be fixed on the actual spectrum. On the other hand, the model contains free phenomenological parameters, namely the mixing angle of $K_1$ states, $\theta_{K_1}$, and the quark-pair-creation constant $\gamma$. These parameters are to be adjusted on the strong decay experiments themselves (additional information on the mixing can be obtained from the mass spectrum, or other types of decays).
\vspace*{0.4cm}

\noindent
4 \underline{\it  The plan of the paper}\\
In Section~\ref{sec:K1_overview}, we present a brief summary of the present status of the experience concerning the $K_1$-mesons. In Section~\ref{sec:th_model}, after having discussed the basic question of the mixing, we introduce the formalism of the theoretical model, namely the $^3P_0$ quark-pair-creation model, used to predict the partial wave amplitudes for the quasi-two-body decays of the $K_1$-meson. In Section~\ref{sec:comparison_th-exp}, we establish the general relation between our model predictions and the most extensive experimental results obtained by ACCMOR collaboration, which use the $K$-matrix formalism to analyse the partial waves. We describe some of the problems we have observed, which include the definition of the total $K_1$-width, the phase space and threshold effects, the strong phase between different intermediate resonance ("isobar") states. In Section~\ref{sec:results}, combining the experimental results on the $K_1$-decays and the predictions of the quark-pair-creation model, we determine the phenomenological parameters of this decay model, the $K_1$ mixing angle $\theta_{K_1}$ and the universal quark-pair-creation constant $\gamma$, and we present the resulting numerical predictions. We compare our model predictions and the measurements of the ACCMOR and Belle collaborations. We also discuss the issues of the relative strong ``offset'' phases and the controversial $K_1(1270)\to\kappa\pi$ channel. We give our conclusions and perspectives in Section~\ref{sec:conclusions}.

\section{Overview of the previous experimental $K_1$-decay studies}\label{sec:K1_overview}

Here we summarise the experimental results of the axial vector $K_1$-resonance study.

\begin{enumerate}
	\item Two close in mass axial-vector mesons, $K_1(1270)$ and $K_1(1400)$, were disentangled in the experiments on the diffractive production of the $1^+(K\pi\pi)$ system in the $Kp\to K\pi\pi p$ reaction, first by the group at SLAC~\cite{Carnegie:1977uz} and then by the ACCMOR collaboration in WA3 experiment at CERN~\cite{Daum:1981hb}. They also observed separately: one $K_1(1270)$ in the strangeness-exchange reaction $\pi^-p\to\Lambda K\pi\pi$~\cite{Rodeback:1980zt} and the other $K_1(1400)$ in the charge-exchange reaction $K^-p\to\overline{K}^0\pi^+\pi^-n$~\cite{Aston:1986jb}. In our study we rely mainly on the diffractive reactions which allow a more detailed study. The relative ratios of two dominant channels, $K^*\pi$ and $K\rho$, indicate that $K_1(1400)$ decouples from the $K\rho$, while the $K\rho$ decay mode of $K_1(1270)$ is dominant (see Table~\ref{tab:BRK1_Daum}). This decay pattern suggests that the observed mass eigenstates, $K_1(1270)$ and $K_1(1400)$, are the mixtures of two strange axial-vector $SU(3)$ octet states $K_{1A}(^3P_1)$ and $K_{1B}(^1P_1)$, as explained later.
	\item The $K_1$-resonances were also observed and studied in $\tau$-decays, $\tau\to K_1\nu_\tau$, by TPC/Two-gamma~\cite{Bauer:1993wn}, ALEPH~\cite{Barate:1999hj}, OPAL~\cite{Abbiendi:1999cq} and CLEO~\cite{Asner:2000nx} collaborations. 
	\item Radiative $B$-decays involving the $K_1$-mesons were also observed by the Belle collaboration~\cite{Yang:2004as}. The data indicate that $\mathcal{B}(B\to K_1(1270)\gamma)\gg\mathcal{B}(B\to K_1(1400)\gamma)$.	
	\item Quite recently the Belle collaboration published a paper on $B\to J/\psi(\psi^\prime)K\pi\pi$ decays~\cite{Belle:2010if}, which will be discussed in detail later.
	\item In addition, the BABAR collaboration reported the measurement of the branching ratios of neutral and charged $B$-meson decays to final states containing a $K_1(1270)$ and $K_1(1400)$-meson and a charged pion: $B^0\to K_1(1270/1400)^+\pi^-$ and $B^+\to K_1(1270/1400)^0\pi^+$~\cite{Babar:2009ii}. In order to parametrize the signal component for the production of the $K_1$-resonances in $B$-decays, the $K$-matrix formalism, used in the analysis by Daum {\it et al.} in~\cite{Daum:1981hb}, was applied for the model description. Since only some parameters, used in the analysis of the ACCMOR collaboration, have been reported, the BABAR collaboration refitted the ACCMOR data in order to determine the parameters describing the diffractive production of the $K_1$-mesons and their decays. One observes that  some results are somewhat different. 
\end{enumerate}

\begin{table}[h]\centering
	\begin{tabular}{|c|c|c|c|c|c|}
		\hline
		$K_1$ & \begin{tabular}{c}$M_{K_1}^{ACCMOR}$,\\ GeV/$c^2$\end{tabular} & \begin{tabular}{c}$\Gamma_{K_1}^{ACCMOR}$,\\ MeV/$c^2$\end{tabular} & $\mathcal{B}(K^*\pi)_S$ & $\mathcal{B}(K^*\pi)_D$ & $\mathcal{B}(K\rho)_S$ \\
		\hline
		$K_1(1270)$ & 1.27$\pm$0.007 & 90$\pm$8 & 0.13$\pm$0.03 & 0.07$\pm$0.006 & 0.39$\pm$0.04 \\
		\hline
		$K_1(1400)$ & 1.41$\pm$0.025 & 165$\pm$35 & 0.87$\pm$0.05 & 0.03$\pm$0.005 & 0.05$\pm$0.04 \\
		\hline
	\end{tabular}		
	\caption{Fitted masses, total widths and partial branching ratios of $K_1(1^+)$ decays into vector-pseudoscalar states, measured by the ACCMOR collaboration in the $Kp\to K\pi\pi p$ reaction for the low momentum transfer to the recoiling proton~\cite{Daum:1981hb}, and tabulated in PDG. The total widths seem to be misleading for the calculation of partial widths, as discussed later in the text.}
	\label{tab:BRK1_Daum}
\end{table}

\section{The theoretical model}\label{sec:th_model}

Before presenting  the $^3P_0$, we begin by explaining the question of the mixing of $K_1$ sates, which is a basic assumption of all the approaches, since there is no theoretical approach predicting quantitatively this mixing.

\subsection{The mixing of the $K_1$ resonances}

In the quark model there are two possible states for the orbitally excited axial-vector mesons: $J^{PC}=1^{++}$ and $J^{PC}=1^{+-}$, depending on different spin couplings of two constituent quarks. In the $SU(3)$-limit these states do not mix in general, but since the $s$-quark is actually heavier than the $u$- and $d$-quarks, the observed $K_1(1270)$- and $K_1(1400)$-mesons are not pure $1^3P_1$ or $1^1P_1$ states. They are considered to be mixtures of non mass eigenstates $K_{1A}$ and $K_{1B}$. Introducing a $K_{1A}-K_{1B}$ mixing angle $\theta_{K_1}$, mass eigenstates can be defined in the following way~\cite{Suzuki:1993yc}~\footnote{To be able to compare with other mixing angle estimations, one has to be careful due to the different parametrizations that are used in the literature. For instance, in the analysis by Carnegie {\it et al.}~\cite{Carnegie:1977uz} the parametrization is $|K_1(1270)\rangle=|K_{1A}\rangle\cos\theta_{K_1}^{(SLAC)}+|K_{1B}\rangle\sin\theta_{K_1}^{(SLAC)}$, $|K_1(1400)\rangle=-|K_{1A}\rangle\sin\theta_{K_1}^{(SLAC)}+|K_{1B}\rangle\cos\theta_{K_1}^{(SLAC)}$. To compare with the results made by Daum {\it et al.}~\cite{Daum:1981hb}, parametrization is written as follows: $|K_1(1270)\rangle=-|K_{1A}\rangle\sin\theta_{K_1}^{(ACCMOR)}+|K_{1B}\rangle\cos\theta_{K_1}^{(ACCMOR)}$, $|K_1(1400)\rangle=|K_{1A}\rangle\cos\theta_{K_1}^{(ACCMOR)}+|K_{1B}\rangle\sin\theta_{K_1}^{(ACCMOR)}$. Comparing the fitted effective couplings one can see that the coupling to $K_{1B}$ has a different sign in these two definitions. Since one can measure only the absolute value of the amplitude, this sign changes nothing and hence it is possible to redefine the sign of this coupling in the paper by Daum {\it et al.}. After that one can easily establish the correspondence between these two forms of parametrization and the one we use in this paper: $\theta_{K_1}=\theta_{K_1}^{(ACCMOR)}=90^\circ-\theta_{K_1}^{(SLAC)}$.}:

\begin{equation}
	\begin{split}
	|K_1(1270)\rangle &= |K_{1A}\rangle\sin\theta_{K_1}+|K_{1B}\rangle\cos\theta_{K_1} \\
	|K_1(1400)\rangle &= |K_{1A}\rangle\cos\theta_{K_1}-|K_{1B}\rangle\sin\theta_{K_1}
	\label{eq:K1_mixing}
	\end{split}
\end{equation}

Since all of $SU(3)$ operators can be expressed as combinations of isospin, $U$- and $V$-spin operators, if an operator describing the interaction is invariant under the $SU(3)$-group transformations, it is also invariant under the isospin, $U$-spin and $V$-spin transformations~\cite{Lipkinbook}. However, it is sufficient to require the invariance only under the isospin and $U$-spin (or $V$-spin) transformations, since $V$-spin is dependent on the isospin and $U$-spin and the $V$-spin operators can be obtained from the $U$-spin operators by an isospin transformation ($U$-spin can be turned into $V$-spin via rotation by 120$^\circ$). 
 
Analogously to $G$-parity, one can define $U$- and $V$-parities: $G_U=C(-1)^U$ and $G_V=C(-1)^V$ respectively, where $C$ is the charge-conjugation parity of the neutral non-strange members of the multiplet. The neutral and charged kaons in the octets are the eigenstates of $U$- and $V$-parities and always have $U$ or $V=1$ respectively.

In the $SU(3)$-limit two kaons that belong to the octets of the same spin but opposite $C$-parity can not mix. To illustrate it, one can consider a matrix element of some arbitrary operator $\mathcal{O}$ between two neutral kaons from different octets ~\cite{Kane:1967xx,Kane:1968zz}:

 \begin{equation}
	 \langle K_A|\mathcal{O}|K_B\rangle = \langle K_A|G_U^{-1}G_U\mathcal{O}G_U^{-1}G_U|K_B\rangle = C_AC_B\langle K_A|G_U\mathcal{O}G_U^{-1}|K_B\rangle
	 \label{}
 \end{equation}
 If the $\mathcal{O}$ operator is $SU(3)$-invariant, i.e. $G_U\mathcal{O}G_U^{-1}=\mathcal{O}$, the matrix element of the transition $\langle K_A|\mathcal{O}|K_B\rangle=0$ unless $C_A=C_B$. 

Strong interactions can break the $SU(3)$-symmetry and produce the mass splittings. It is experimentally confirmed that isospin is conserved in strong interactions. Hence, if the strong interaction operator breaks the $SU(3)$-symmetry, $U$- and $V$-parities are not conserved anymore, even if $G$-parity is conserved. In this case $G_U\mathcal{O}G_U^{-1}\neq\mathcal{O}$ and consequently $\langle K_A|\mathcal{O}|K_B\rangle\neq0$ and the mixing takes place.

That this mixing is indeed the effect of the symmetry breaking can be explicitly seen in quark models. At the level of bound states of a potential model. It is induced, for instance, by spin-orbit forces with different $s$ and $u,d$ quark masses. A mixing is also generated by the two-meson loops, due to quark pair creation and annihilation in the bound states, as is explained below in the $K$-matrix approach, subsection \ref{K-MQ}. In this approach, the real mixing must be understood as the one of the $K$-matrix couplings corresponds to the effect of the real part of the loops, while additional, complex, mixing would be present in the physical couplings.
 
One can see easily why the loops and the $SU(3)$ breaking generate mixing. For instance, the $K^*\pi$ and $K\rho$ loop contributions connect the $K_A$ and the $K_B$, since both states are coupled to these channels. In this way it generates the mixing. The two contributions cancel each other if one sets $M_{K^*}=M_{\rho}$ and $m_\pi=m_K$, i.e. in the case of the exact $SU(3)$-symmetry. It must be emphasized however that this mechanism of loops does not lead to the actual calculation of the mixing angle, because one would have to sum over a very large number of possible intermediate states. Therefore, in this approach, it remains an independent phenomenological parameter, which has to be fixed through confrontation with data. 

It is to be noted that, anyway, no fundamental calculation of the mixing has been produced.

\subsubsection{Previous phenomenological determinations of the mixing angle}\label{sec:mxing-det.}

Here, we gather all the various estimations of $\theta_{K_1}$.

On the other hand, there have been, in the past, many attempts to determine the mixing angle, both from experimentalists and theoreticians, but in both cases only through phenomenological analyses. The phenomenological analyses have concerned the masses (with additional assumptions, since $SU(3)$ alone does not enable to fix the mixing angle from the masses), the $\tau\to K_1\nu_\tau$ decays, the $B\to K_1$ transitions, and, mainly, the strong decays $K_1\to K\pi\pi$ through 
$K^*\pi$ and $K\rho$ channels. Indeed, the pattern of the latter is very sensitive to the mixing angle. 

The angles are given according to the definition above, Eq.~\eqref{eq:K1_mixing}. However, one must warn that it does not completely fix the definition, since there may be different choices of the phases of the states. In general, it is difficult to establish completely the connection with our own definition in the present paper, so we only state the absolute magnitude of the angle.

		\begin{itemize}
			\item In the experiment, carried out at SLAC by Carnegie {\it et al.}~\cite{Carnegie:1977uz}, the mixing angle was determined from the $SU(3)$ couplings to the $K^*\pi$ and $K\rho$ channels to be $\theta_{K_1}=(41\pm4)^\circ$. On the other hand, the partial wave analysis of the WA3 experiment data, done by the ACCMOR collaboration (Daum {\it et al.} \cite{Daum:1981hb}), gives $\theta_{K_1}=(64\pm8)^\circ$ and $\theta_{K_1}=(54\pm4)^\circ$ for the low and high momentum transfer to the recoiling proton respectively.

			\item  In the reanalysis of ACCMOR data by BABAR\cite{Babar:2009ii}, using the low $t$-data, the refitted value of the $K_1$ mixing angle turns out to be $72^\circ$ compared to $64^\circ$ from the ACCMOR fit.

			\item In the work by Suzuki~\cite{Suzuki:1993yc}, the mixing angle is determined by three different approaches. One is in order to explain the observed hierarchy in the $K_1$ strong decays to $K^*\pi$ and $K\rho$, like has been done by SLAC and ACCMOR. Another is the $SU(3)$ analysis of the masses of the two octets, but with additional assumptions . Finally, the suppression of $\tau \to K_1 (1400) \nu$ with respect to $\tau \to K_1 (1270) \nu$ is considered. Two possible solutions for the $K_1$ mixing angle were found: $\theta_{K_1}\approx33^\circ$ or $57^\circ$.

			\item In the analysis of $\tau \to K_1 \nu$ done by the CLEO collaboration~\cite{Asner:2000nx}, the $K_1$ mixing angle was determined from the measured ratio $\mathcal{B}(\tau\to K_1(1270)\nu_\tau)/\mathcal{B}(\tau\to K_1(1400)\nu_\tau)$: $\theta_{K_1}=(69\pm16\pm19)^\circ$ for $\delta=0.18$ and $\theta_{K_1}=(49\pm16\pm19)^\circ$ for $\delta=-0.18$ where $|\delta|=\frac{m_s-m_u}{\sqrt2(m_s+m_u)}\approx0.18$ is a phenomenological $SU(3)$ breaking parameter. This result is consistent with the calculation by Suzuki~\cite{Suzuki:1993yc}.
 %The latest measurements \cite{Barate:1999hj,Abbiendi:1999cq,Asner:2000nx} show that the $K_1(1270)$ production is favoured over the $K_1(1400)$ production: $\mathcal{B}(\tau^-\to K_1(1270)^-\nu_\tau)=(4.7\pm1.1)\times10^{-3}$ while $\mathcal{B}(\tau^-\to K_1(1400)^-\nu_\tau)=(1.7\pm2.6)\times10^{-3}$~\cite{PDG}.

			\item In the work of Blundell, Godfrey and Phelps~\cite{Blundell:1995au}
			
\begin{itemize}
			\item The mixing is discussed using the results of the TPC/Two-gamma collaboration: $\mathcal{B}(\tau^-\to K_1(1270)^-\nu_\tau)=(0.41^{+0.41}_{-0.35})\%$ and $\mathcal{B}(\tau^-\to K_1(1400)^-\nu_\tau)=(0.76^{+0.40}_{-0.33})\%$. This would seem to mean that the rate into $K_1(1400)$ is larger than into $K_1(1270)$, although their errors are too large to make a strong statement. Anyway these numbers have been superseded by the CLEO data, which show the contrary.

			\item The strong decays of the $K_1$-mesons to the final states $K^*\pi$ and $K\rho$ were studied as well in order to determine the mixing angle. A $\chi^2$ fit of the experimental data on the partial decay widths $\Gamma(K_1(1270/1400)\to K^*\pi)$ and $\Gamma(K_1(1270/1400)\to K\rho)$ was used for the $\theta_{K_1}$-determination.

\begin{itemize}
					\item Performing a $\chi^2$-fit with the predicted decay widths, calculated within the pseudo-scalar-meson-emission model, using simple harmonic oscillator wave functions with a single parameter $\beta=0.40$~GeV, the fitted value of the mixing angle was obtained to be $\theta_{K_1}=(48\pm5)^\circ$.
					\item The strong $K_1$-decays were also calculated using both the flux-tube-breaking model and the $^3P_0$ model for several sets of meson wave functions. In all cases a second fit was performed by allowing both $\theta_{K_1}$ and the quark-pair-creation constant $\gamma$ to vary, what reduces the $\chi^2$ significantly. Using simple harmonic oscillator wave functions with $\beta=0.40$~GeV, comparison of the predicted decay widths by the $^3P_0$ model to experimental results gives $\theta_{K_1}=(45\pm4)^\circ$, while the flux-tube-breaking model's prediction gives $\theta_{K_1}=(44\pm4)^\circ$, both appreciably different from our central value $\theta_{K_1}\simeq 60^\circ$ with the same set of wave functions. Their last result for $\theta_{K_1}$ is slightly changed for the case of use of different set of the meson wave functions from Ref.~\cite{Godfrey:1985xj}: $\theta_{K_1}=(51\pm3)^\circ$.
				\end{itemize}
		\end{itemize}

	  \item A detailed study of the $B\to K_1(1270)\gamma$ and $B\to K_1(1400)\gamma$ decays in the light-cone QCD sum rules approach was presented by Hatanaka and Yang in~\cite{Hatanaka:2008xj}.  The sign ambiguity of the mixing angle is resolved by defining the signs of the decay constants $f_{K_{1A}}$ and $f_{K_{1B}}^\perp$.
\begin{itemize}
		\item From the comparison of the theoretical calculation and the data for decays $B\to K_1\gamma$ and $\tau\to K_1\nu_\tau$, it was found that $\theta_{K_1}=-(34\pm13)^\circ$ is favoured within the conventions of Hatanaka and Yang. It is difficult to establish the relation with our own convention as regards sign.
		\item The predicted branching ratios, $\mathcal{B}(B\to K_1(1400)\gamma)$ and $\mathcal{B}(B\to K_1(1270)\gamma)$, are then in agreement with the Belle collaboration measurement within the errors. 
\end{itemize}		
		\end{itemize}
\vskip 5mm
In summary, the cleanest way to extract the mixing angle is certainly, in principle, the determination from the ratio of $\mathcal{B} (\tau\to K_1(1400/1270)\nu_\tau)$, if the data were sufficiently accurate. At present, we believe that the best way remains the study of strong decays, as we do in this paper.

\subsection{The $^3P_0$ Quark-Pair-Creation Model}\label{sec:QPCM}

There are several additive quark models of strong vertices. All these models relate to the recoupling coefficients of unitary spin, quark spin and the quark orbital angular momenta, but differ in the dynamical description. One of the simplest additive quark model describing three-meson vertices is the naive quark-pair-creation model (QPCM), with a $^3P_0$ structure for the pair, formulated by Le~Yaouanc, Oliver, P\`ene and Raynal~\cite{LeYaouanc:1972ae} starting from ideas of Micu and of Carlitz and Kislinger~\cite{Micu:1968mk,Carlitz:1970xb}. The model has then been extensively applied and discussed by many authors, including the same authors (see Ref.~\cite{Gavela:1978bz} and some references therein) and the group of N.~Isgur in Canada~(for instance Refs.~\cite{Kokoski:1985is,Blundell:1995ev}). As in the usual additive quark models with spectator quarks, the quark-antiquark pair is ``naively'' created not from the ingoing quark lines but within the hadronic vacuum. The strong interactions vertices in the QPCM are expressed in terms of the explicit harmonic oscillator spacial $SU(6)$ wave functions (compared to the work by Micu~\cite{Micu:1968mk}, who just fitted the various spacial integrals using the measured decay widths, what does not allow to study the polarization effects) and a nonlocal vacuum quark-aniquark pair production matrix element, depending on the internal quark momenta (while Carlitz and Kislinger~\cite{Carlitz:1970xb} neglected the internal momentum distributions). Contrary to the QPCM by Colglazier and Rosner~\cite{Colglazier:1970vx}, the $^3P_0$ structure of the created pair describes any decay process of any hadron, using one universal parameter. The other model parameters are those of the hadrons themselves (potential model), and not relative to the decay process as in~\cite{Colglazier:1970vx}, where the various extra couplings between the pair and the incoming meson depend on the nature of the hadron states and may be weighted by different arbitrary coefficients for different hadrons. 

The naive QPCM has the advantage of making definite predictions for all hadronic vertices and moreover, contrary to the other works, it predicts the relative signs of the couplings. Another appealing feature of the model is that it consists only one phenomenological parameter (the quark-pair-creation constant), what allows a much more general description and relates the amplitudes of different processes. The main weakness of the QPCM is that the emitted hadrons are considered to be non-relativistic. Thus one has to look for the decays that are not significantly sensitive to these effects.

A specific study of the strange axial-vector mesons was first done by Blundell, Godfrey and Phelps~\cite{Blundell:1995au}, who studied the properties of $K_1$ by combining wave functions inspired by the Godfrey-Isgur quark model to describe the bound states, and the flux-tube-breaking or $^3P_0$ models to describe the decays. Although we start from the same basic $^3P_0$ model, we give a much more extended study, which is, in particular, required for the purpose of the $\lambda_\gamma$ determination. We make a rather different discussion, especially, for the relation between theory and experiment. We clarify the relation with the $K$-matrix analysis, which is the tool used by the main experiment, that is the ACCMOR experiment. We discuss the definition of widths, which appears very ambiguous due to threshold effects.  We also include a treatment of the off-shellness (i.e. damping factor). 
In addition, we explore the system of phases, which is one main achievement of the model (as well as it has been in the baryon decays). Finally, we discuss in detail the most problematic $\kappa\pi$ channel. These differences will become  apparent from the rest of the paper.

\subsubsection{Formalism}

\begin{figure}[h!]\centering
	\includegraphics[width=0.45\textwidth]{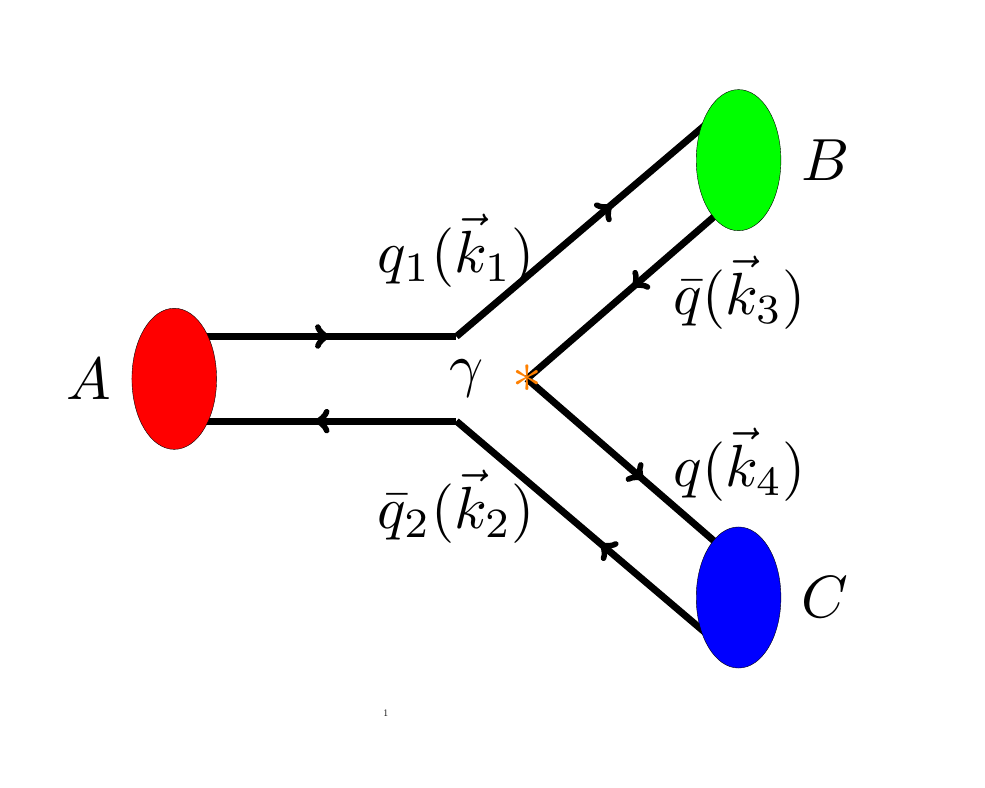}
	\caption{Three-meson vertex in the quark-pair-creation model.}
	\label{fig:QPCpic}
\end{figure}

In the QPCM, instead of being produced from the gluon emission, the quark-antiquark pair $q\bar{q}$ (see Fig.~\ref{fig:QPCpic}) is created anywhere within the hadronic vacuum by an operator proportional to $(u\bar{u}+d\bar{d}+s\bar{s})~{\bf S}\cdot{\bf p}$ where ${\bf S}$ refers to spin 1 and ${\bf p}$ is the relative momentum of the pair. It is combined with the initial quark-antiquark system $\bar{q}_2q_1$ and produces the final state $B(q_1\bar{q})C(q\bar{q}_2)$. The initial spectator quarks are supposed not to change their $SU(3)$ quantum numbers, nor their momentum and spin. In order to conserve the vacuum quantum numbers the pair must be created in the $^3P_0$ state due to $P=-(-1)^L$ and $C=(-1)^{L+S}$ parity conservation with 0-total momentum ($\vec{k}_3+\vec{k}_4=0$) and to be a $SU(3)$-singlet. Thus the matrix element of the quark-antiquark pair production from the vacuum is unambiguously constructed with the help of the spins and momenta of the quark and antiquark only~\cite{LeYaouanc:1972ae}:

\begin{equation}
	\langle \bar{q}q|\hat{T}_{vac}|0\rangle = \delta(\vec{k}_3+\vec{k}_4)\gamma\sum_m(1,m;1,-m|0,0)\mathcal{Y}_1^m(\vec{k}_3-\vec{k}_4)\chi_1^{-m}\phi_0
	\label{}
\end{equation}
where $\gamma$ is a phenomenological dimensionless pair-creation constant (which is determined from the measured partial decay widths and taken to be of the order of 3-5), $\chi_1^{-m}$ are the spin-triplet wave functions, $\phi_0=\frac{1}{\sqrt3}(u\bar{u}+d\bar{d}+s\bar{s})$ is the $SU(3)$-singlet and $\mathcal{Y}_1^m$ represents the $L=1$ angular momentum of the pair.

Taking the matrix element of the pair-creation operator between the $SU(6)$ harmonic-oscillator wave functions of hadrons, the matrix element for the decay $A\to B+C$ can be written as:

\begin{equation}
	\langle BC|\hat{T}|A\rangle=\gamma \sum_m(1,m;1,-m|0,0)\Phi_B\Phi_C\Phi_A^m\Phi_{vac}^{-m} I_m^{(ABC)}
	\label{}
\end{equation}
where $\Phi=\chi_1^m\phi$ are the $SU(6)$ spin-flavour wave functions and $I_m^{(ABC)}$ are the spacial integrals dependent on the momentum of the final states, which are computed in Appendix.

Assuming $A$, $B$ and $C$ to be an axial vector, pseudoscalar and vector mesons respectively, the spin part of the matrix element can be written as

\begin{equation}
	\begin{split}
	&\chi_C\chi_B\chi_A\chi_{pair} = \sum_{m_i}(\frac{1}{2},m_1;\frac{1}{2},m_3|0,0) (\frac{1}{2},m_4;\frac{1}{2},m_2|1,\lambda_C) \\
	&\times (\frac{1}{2},m_1;\frac{1}{2},m_2|S_A,m_{S_A}) (1,m_{L_A};S_A,m_{S_A}|1,\lambda_A) (\frac{1}{2},m_4;\frac{1}{2},m_3|1,-m)
	\end{split}
\end{equation}

Consider for instance $K^{*0}\pi^+$ decay mode of $K_1$-meson. After the summation over the spin projections the calculated helicity amplitudes for the $K_{1A}$ ($1^3P_1$) and $K_{1B}$ ($1^1P_1$) will be (the definition of the helicity amplitudes and their relation with the partial wave amplitudes can be found in Appendix):

\begin{equation}
	\begin{split}
		&\mathcal{M}_{00}^{10(A)} = -\gamma\frac{I_1^{(K_1K^*\pi)}}{3\sqrt2}, ~ ~ ~ ~ ~ ~ ~\mathcal{M}_{10}^{11(A)} = -\gamma\frac{I_1^{(K_1K^*\pi)}-I_0^{(K_1K^*\pi)}}{6\sqrt2} \\
	&\mathcal{M}_{00}^{10(B)} = -\gamma\frac{I_0^{(K_1K^*\pi)}}{6}, ~ ~ ~ ~ ~ ~ ~\mathcal{M}_{10}^{11(B)} = \gamma\frac{I_1^{(K_1K^*\pi)}}{6}
	\end{split}
	\label{K1ABhelicity}
\end{equation}
The corresponding amplitudes for the $K^+\rho^0$ mode are obtained by multiplying the $K^{*0}\pi^+$ amplitudes by $1/\sqrt2$ and changing the sign of $K_{1A}$-part. 

Taking into account the isospin factors for different charge states~\footnote{The amplitudes were calculated for $K_1^+\to K^{*0}\pi^+$ and $K_1^+\to K^+\rho^0$. The amplitude of $K\rho$ must be divided over $\sqrt2$ due to isospin wave function of $\rho^0$. To obtain the general amplitude which doesn't depend on the charge combination one has to divide over the isopin factor: $-\sqrt{2/3}$ for $K^*$ and $\sqrt{1/3}$ for $\rho$ since for the matching with the relativistic form factors the charge combination is not relevant. Finally one obtains the factor $\sqrt{3/2}$ in Eq.~\eqref{eq:SDfunctions}.}, the generalized amplitudes are summarized in Table~\ref{tab:K1ABamplitudes}. The functions $S$ and $D$ are defined as

\begin{equation}
	S^{(ABC)}=\gamma\sqrt{\frac{3}{2}}\frac{2I_1^{(ABC)}-I_0^{(ABC)}}{18},~ ~ ~ D^{(ABC)}=\gamma\sqrt{\frac{3}{2}}\frac{I_1^{(ABC)}+I_0^{(ABC)}}{18}
	\label{eq:SDfunctions}
\end{equation}

\begin{table}[h]\centering
	\begin{tabular}{|c|c|c|}
		\hline
		Decay mode & $A_S$ & $A_D$ \\
		\hline
		$K_{1B}\to K^*\pi$ & $-S^{(K_1K^*\pi)}$ & $-\sqrt2D^{(K_1K^*\pi)}$ \\
		\hline
		$K_{1A}\to K^*\pi$ & $\sqrt2S^{(K_1K^*\pi)}$ & $-D^{(K_1K^*\pi)}$ \\
		\hline
		$K_{1B}\to K\rho$ & $S^{(K_1K\rho)}$ & $\sqrt2D^{(K_1K\rho)}$ \\
		\hline
		$K_{1A}\to K\rho$ & $\sqrt2S^{(K_1K\rho)}$ & $-D^{(K_1K\rho)}$ \\
		\hline
	\end{tabular}
	\caption{Partial wave amplitudes of $K_{1A}(1^3P_1)$ and $K_{1B}(1^1P_1)$ decays into vector-pseudoscalar states, calculated within QPCM.}
	\label{tab:K1ABamplitudes}
\end{table}

One has to point out that our treatment obeys the $SU(3)$-symmetry. $SU(3)$ breaking effects are present only in two places: 1) we use the physical observed masses of hadrons to calculate the momentum transfer of the decay and the phase space; 2) we introduce mixing between the $K_{1A}$ and $K_{1B}$ states.

Then the decay amplitudes of the physical $K_1$ states into $K^*\pi$ or $K\rho$ final states can be expressed as functions of the pseudoscalar meson momentum in the $K_1$ reference frame and the mixing angle $\theta_{K_1}$:

\begin{equation}
	\begin{split}
		A_S(K_1(1270)\to K^*\pi/K\rho) &= S^{(K_1K^*\pi/K_1K\rho)}(\sqrt{2}\sin\theta_{K_1}\mp\cos\theta_{K_1}) \\
		A_D(K_1(1270)\to K^*\pi/K\rho) &= D^{(K_1K^*\pi/K_1K\rho)}(-\sin\theta_{K_1}\mp\sqrt{2}\cos\theta_{K_1}) \\
		A_S(K_1(1400)\to K^*\pi/K\rho) &= S^{(K_1K^*\pi/K_1K\rho)}(\sqrt{2}\cos\theta_{K_1}\pm\sin\theta_{K_1}) \\
		A_D(K_1(1400)\to K^*\pi/K\rho) &= D^{(K_1K^*\pi/K_1K\rho)}(-\cos\theta_{K_1}\pm\sqrt{2}\sin\theta_{K_1})
	\end{split}
	\label{eq:K1SDamplitudes}
\end{equation}

\noindent Correspondingly, the partial decay widths can be determined by using amplitudes squared from the Eqs.~\eqref{eq:K1SDamplitudes} multiplied by the phase space factors:

\begin{equation}
	\Gamma_{S/D}^{QPCM}(K_1\to VP) = 8\pi^2\frac{E_VE_Pk_P}{M_{K_1}}|A_{S/D}(K_1\to VP)|^2.
	\label{eq:partial_width_QPCM}
\end{equation}

Note that all the signs in the expressions for amplitudes have sense only within definite specific conventions. The ones in our work are defined in Appendix \ref{app:QPCM}. On the other hand, the signs of the products of the couplings of the two successive decay processes from the same $K_1$ state, that is, when we multiply by
the decay amplitude of the isobar, make sense and the relative signs are observable because the final state $K \pi \pi$ is the same, and all phase arbitrariness cancels. It is an important feature of the model that \underline{\it it can predict all these observable signs}. As will be seen in subsection \ref{sec:offset_phases}, these predictions are remarkably verified by experimental data.

\subsubsection{The choice of the spatial wave functions}

The unknown parameters of the model are the quark-pair-creation constant $\gamma$ and the $K_1$ mixing angle, which we determine by fitting the experimental data on the $K_1$-decays (see the next section). However, before proceeding to this determination, the model must be specified by the choice of the set of meson wave functions. In accordance with a fact that the $^3P_0$ model is a simple model, we will remain within the traditional $SU(6)$ approximation which describes rather well ordinary radiative decays (e.g. $\omega\to\pi\gamma$). This includes the $SU(3)$-symmetry approximation which anyway is also present in the $^3P_0$ model through the fact that the quark-pair-creation constant is the same for all reactions. In this approach the effect of the $SU(3)$ breaking is taken into account only through the dependence of the decay momentum of the physical hadronic masses. For practical reasons, we choose a set of harmonic oscillator wave functions, which are known to give a reasonable approximation.

Here one has to stress that the harmonic oscillator radius of the meson wave function ($\psi(r)\propto\exp(-r^2/2R^2)$, for details see Appendix \ref{app:QPCM}) is not a free phenomenological parameter. In principle, it can be predicted by the quark-potential model describing the bound states of two quarks. To get a first and rough estimate we can use the following relation, obtained in the non-relativistic harmonic oscillator model for the energy shift between the ground state and the first radial excitation:

\begin{equation}
	\Delta E_1=\frac{2}{m_qR^2}
	\label{eq:DeltaE1}
\end{equation}
with $m_q$ being the quark mass, which can be standardly estimated from the magnetic moment of the proton: $\mu_p=\frac{e}{2m_q}=\frac{2.79}{2m_N}$. Whence $m_q\simeq0.34$~GeV. $\Delta E_1$ can be estimated from the energy of the $L=1$ state of the order of (1.2-1.3)~GeV and the weighted average energy of the ground state $(3m_\rho+m_\pi)/4\simeq0.6$~GeV. Then the estimated radius is given by

\begin{equation}
	R=\sqrt{\frac{2}{\Delta E_1 m_q}}\sim \sqrt{\frac{2}{(1.25-0.6)~0.34}}\simeq 3~\text{GeV}^{-1}
	\label{}
\end{equation}

On the other hand, it is obvious that this approximation of the Schroedinger equation with the harmonic oscillator potential is rather naive: the realistic potential is known to be of the form of linear (that describes confinement) plus Coulomb potential. One has also to notice that the application of the use of the non-relativistic character of the Schroedinger equation to the heavy-light systems is dubious. Therefore, one could take a value inspired by the well known model of Godfrey and Isgur. Of course, in the latter model the solutions are no longer the harmonic oscillator wave functions. However, such harmonic oscillator wave functions can represent a good approximation if the radius $R$ is adjusted. For most $L=0,1$ states one finds in this model the typical value $R\sim2.5~\text{GeV}^{-1}$~\cite{Kokoski:1985is}. For our predictions we therefore adopt a set of wave functions with a \underline{\it common} harmonic oscillator radius having precisely this value,

\begin{equation}
	R=2.5~\text{GeV}^{-1}
	\label{R25}
\end{equation}

This is one of the choices made by Blundell {\it et al.}~\cite{Blundell:1995au}. We must warn that in the model of Godfrey and Isgur, pion and kaon have actually quite smaller radius ($\sim1.4~\text{GeV}^{-1}$ \cite{Kokoski:1985is}) due to the strong spin-spin interaction force.  If we were adopting the low values for the Goldstone boson we would obtain unsatisfactory results. For example, using $R_\pi\simeq1.4~\text{GeV}^{-1}$, we can not reproduce correctly the $D/S$ ratio in the $b_1\to\omega\pi$ decay which is precisely measured ($D/S=0.28$). The use of the exact wave functions of the model of Godfrey and Isgur \cite{Godfrey:1985xj} does not seem to improve the situation; one finds $D/S=2.5/14\simeq0.18$ from the tables of Kokoski and Isgur \cite{Kokoski:1985is}.

Of course, although it has not been commented in previous works, this fact is disturbing, since the spin-spin force is present indeed in spectrocopy, and therefore, it should be more realistic to include its effect. In addition to empirical success, the choice of equal radii can be motivated in the spirit of the $SU(6)$ approach. It must be remembered  indeed that old quark model very naive calculations have succeeded well with this $SU(6)$ symmetry, for instance to relate $\omega\to\pi\gamma$ to magnetic moments. Now, the $^3P_0$ model is also in this very naive spirit: it is non-relativistic in essence.

\subsection{The issue of the damping factor}

In the end of this introduction of the theoretical model, we discuss the necessity of introducing an additional cut-off on momenta (or damping factor) in the coupling vertices. Generally speaking, there is need of a strong  cut-off for calculations involving far off-shell particles, once the model has been adjusted on real decays. Indeed, the natural fall-off provided by simple continuation of the $^3P_0$ model, due to the wave functions, is seen to be much too weak. The need for this cutoff appears in various circumstances:

\begin{itemize}
	\item In the \underline{\it branching ratios}, obtained by the integration over a large phase space, like for the production of $K\pi\pi$ (e.g. $B\to K\pi\pi\gamma,\psi$) or similar. For instance, Belle \cite{Belle:2010if} defines branching ratios by the ratios of integrals over the whole phase space. If there were not such a cutoff, the higher partial wave contribution like $D$-waves would be found much too large with respect to $S$ waves, due to the centrifugal barrier factors  $k^{2 l}$, which increase too much at large  mass of the $K\pi\pi$ system~\footnote{Experimentally, the problem does not appear in the work of Belle, because they do not introduce $D$-waves for the $K_1$-decays.}.

	\item Departure of the resonance line shape from the Breit-Wigner formula.
	Resonances are usually described by multiplying the standard Breit-Wigner (and the width) by the so-called ``centrifugal barrier'' factors. The term is ambiguous, since these factors includes both the proper centrifugal barrier effect, which is the universal $k^{2l}$ automatically present in partial waves (increasing with the momentum), and the damping factor, which is highly model dependent, and decreases with the momentum.
	In fact the prototype of such factors are the Blatt-Weisskopf factors of nuclear physics, also commonly used by experimentalists in particle physics. They are deduced for a spherical well potential, which is obviously very naive. One consequence of this particular set of factors is that there would be no damping for $S$ waves, which is not true in more realistic models
	(harmonic oscillator wave functions in the $^3P_0$ model give a Gaussian damping in all waves).

	 The accurate studies of the resonance shapes show directly a departure from the standard Breit-Wigner shape, e.g. for the $\Delta(1236)$~\cite{Angela} or the $K^*(890)$, see Ref.~\cite{Aston:1986jb}).

	\item \underline{\it Contribution of loops to the self-energy}

	The need for the cutoff  is also shown by calculations of the hadronic loop contribution to the self-energy of mesons (see subsection \ref{sec:relation-K-MQ}), which involves integration over the possible momentum up to infinity. In the $^3P_0$ model~\cite{SilvestreBrac:1991pw}, the contribution to the self-energy would be much too large for $D$-waves, in spite of the cutoff naturally provided by the gaussian wave functions, yielding finally a bad spectrum. 

One obtains a natural damping factor through the Gaussian factors $e^{-\beta k^2}$:

\begin{equation}
	A_S\propto (3-\alpha k^2)e^{-\beta k^2}, ~ ~ ~A_D\propto \alpha k^2 e^{-\beta k^2}
	\label{}
\end{equation}
but one finds $\beta\sim0.3$~GeV$^{-2}$ which is much too small ; it does not reduce efficiently the $D$ waves contributions for the loops and neither for the off-shell situations we consider. Following Ref.~\cite{SilvestreBrac:1991pw}, we introduce the empirical Gaussian cutoff $\exp[-\beta^\prime(k^2-k_0^2)]$ with $\beta'\approx 3$~GeV$^{-2}$, where $k_0$ is the decay momentum when all the particles are put on-shell:

\begin{equation}
	A_S\propto (3-\alpha k^2)e^{-\beta k^2}\times e^{-\beta^\prime(k^2-k_0^2)}, ~ ~ ~A_D\propto \alpha k^2 e^{-\beta k^2}\times e^{-\beta^\prime(k^2-k_0^2)}
	\label{}
\end{equation}
\end{itemize}

\vskip 0.5cm
With this additional damping factor one finds that the integrated $D/S$-ratio becomes stable. The isobar ($K^*/\rho$) decay does not depend much on the damping factor. However, another effect then appears in the decay rate from the parent $K_1$ to one isobar and one stable particle: integrating over the mass of the isobar, the calculated partial width depends on the presence of the damping factor for the decay of the parent $K_1$ resonance to an off-shell isobar. The low end of the isobar mass spectrum corresponds indeed to large off-shell momenta. This effect has been duely taken into account in our calculations.
\vskip 0.5cm

In the calculation of $\lambda_\gamma$ presented in our previous paper \cite{Kou:2010kn}. the effect of the  introduction of this damping factor in the decay amplitude of the $K_1$ is important. Indeed, the interference of several channels needed to obtain a non-zero imaginary part of $\vec{n}\cdot(\vec{\mathcal{J}}\times\vec{\mathcal{J}}^*)$  requires a large off-shellness of the intermediate isobars. We find that this quantity is sensitive to the presence of the $D$ waves, and then to the introduction of the damping factor.

\section{How to compare the theoretical model computation with the experimental data?}\label{sec:comparison_th-exp}

Let us stress that the use of  experimental data in our work is twofold: first determine the model parameters $\gamma$ and $\theta_{K_1}$,  and then check the validity of our model predictions.

In this section we will explain how one can relate the quark model predictions for the decay partial widths of $K_1$ to the $K$-matrix analysis of Daum {\it et al.}, which is the main source of experimental information. 

Indeed, the main experiments on the $K_1$-decays~\cite{Daum:1981hb,Carnegie:1977uz} were analysed with the same $K$-matrix formalism developed by Bowler {\it et al.}~\cite{Bowler:1974th} and obtained very similar results. We use in our analysis the parameters of the analysis done by Daum {\it et al.}(ACCMOR experiment) which seems to be the most detailed. On the other hand, there are certain physical parameters of the fit which are not tabulated in the this paper. Then we also use, where necessary, the results of the $K$-matrix re-analysis of the ACCMOR data by the BABAR collaboration~\cite{Babar:2009ii}.

Let us now emphasize that the very extensive work of Daum {\it et al.} consists of two distinct steps:

\begin{itemize}
	\item The first one is the partial wave analysis (PWA) where the $K\pi\pi$ three-body final state is decomposed into a sum of quasi-two-body ``partial waves'' ($K^*\pi$, $K\rho$, etc.) with various quantum numbers of the total spin and orbital momentum. In this first step there is no reference to any parent resonance like $K_1$. This step corresponds to the fitted values of the quasi-two-body partial wave amplitudes plotted with the corresponding error bars in \cite{Daum:1981hb}.
	\item The second step is the fit of the partial wave amplitudes, extracted on the previous step, within the $K$-matrix formalism in order to study the structure of the initial parent $K_1$ resonance and its properties (pole masses, couplings to various decay channels, etc.).
\end{itemize}

Let us stress that this two-step procedure is different from the modern Dalitz plot analyses where the isobar and parent resonances are included together in one unique formula of the total amplitude. In that case the total amplitude is written directly as a product of the parent resonance decay amplitude and the amplitude of the subsequent decay of the isobar taking into account the width effects of the unstable resonances by the Breit-Wigner forms. 

We do not question the first step; we rather indicate various difficulties which we have encountered in trying to use the $K$-matrix parameters from the analysis of Daum {\it et al.}. In the following subsection, we first recall the general $K$-matrix formalism and then its relation to the quark model.

\subsection{The $K$-matrix formalism and the quark model}\label{K-MQ}

In order to extract our theoretical parameters, $\gamma$ and $\theta_{K_1}$, we need the experimental partial widths. We also need them to verify our prediction of the model. And the question is: how to define a partial width? Resonances are often parametrized in terms of the Breit-Wigner form

\begin{equation}
	BW_r^{(NR)}(m)\propto \frac{1}{m_r-m-i\frac{\Gamma_r}{2}}, \quad\text{or}\quad BW_r^{(R)}(m)\propto \frac{1}{m_r^2-m^2-im_r\Gamma_r}
	\label{eq:BW}
\end{equation}
in the non-relativistic and relativistic cases respectively. Resonance width, in principle, depends on energy, $\Gamma_r(m)$. This approximation assumes an isolated resonance with a single measured decay. If there is more than one resonance in the same partial wave which strongly overlap, an elegant way that provides the unitarity of the $S$-matrix is to use the $K$-matrix formalism for the two-body decays of the resonance states (for more details see Appendix)~\footnote{Note that in the case of two overlapping resonances the Breit-Wigner parametrization of the amplitude satisfies the unitarity condition of the $S$-matrix only with the complex couplings satisfying certain condition. As we demonstrate later, these complex couplings can be obtained from the real $K$-matrix couplings by a complex rotation.}.

\subsubsection{General definitions in the $K$-matrix formalism}\label{sec:general}

From the unitarity of the $S$-matrix

\begin{equation}
	S\equiv 1+2i\rho^{\frac{1}{2}}T\rho^{\frac{1}{2}}
	\label{eq:S-matrix}
\end{equation}
one gets

\begin{equation}
	T-T^\dagger=2iT^\dagger\rho T=2iT\rho T^\dagger
	\label{eq:Sunitarity1}
\end{equation}
where the diagonal matrix $\rho_{ij}(m)$ is the phase space factor which is discussed in detail later in this section. In terms of the inverse operators Eq.~\eqref{eq:Sunitarity1} can be rewritten as

\begin{equation}
	(T^\dagger)^{-1}-T^{-1}=2i\rho
	\label{}
\end{equation}

\noindent One can further transform this expression into

\begin{equation}
	(T^{-1}+i\rho)^\dagger=T^{-1}+i\rho
	\label{eq:Sunitarity2}
\end{equation}

Using the definition of the $K$-matrix

\begin{equation}
	K^{-1}\equiv T^{-1}+i\rho
	\label{eq:Sunitarity3}
\end{equation}
one can easily find from Eq.~\eqref{eq:Sunitarity2}, \eqref{eq:Sunitarity3} that the $K$-operator is Hermitian, i.e.

\begin{equation}
	K=K^\dagger
	\label{}
\end{equation}

\noindent From the time reversal invariance of $S$ and $T$ it follows that $K$ must be symmetric, i.e. {\it the $K$-matrix can be chosen to be real and symmetric}. Resonances should appear as a sum of poles in the $K$-matrix. In the approximation of resonance dominance one gets therefore

\begin{equation}
	K_{ij}=\sum_{a^\prime}\frac{f_{a^\prime i}f_{a^\prime j}}{m_{a^\prime}-m}
	\label{eq:Kmatrix0}
\end{equation}
where the sum on $a^\prime$ goes over the number of poles with masses $m_{a^\prime}$. In the common approximation in the resonance theory, the couplings $f_{a^\prime i}$ are taken to be real.

The partial and total $K$-matrix widths can be defined as 

\begin{subequations}
	\begin{align}
		\Gamma_{a^\prime i}(m) &= 2f_{a^\prime i}^2\rho_{ii}(m) \\
		\Gamma_{a^\prime}(m) &= \sum_i \Gamma_{a^\prime i}(m)
	\end{align}
	\label{eq:partial_and_total_widths_K}
\end{subequations}

\noindent Note that the $K$-matrix width does not need to be identical with the width which is observed in experiment nor with the width of the $T$-matrix pole in the complex energy plane.

\subsubsection{Relation between the couplings in the $K$-matrix formalism and the quark model}\label{sec:relation-K-MQ}

In this section we identify in a systematic approach the couplings deduced from the $^3P_0$ quark model, including the mixing of $K_1$ resonances, with the couplings introduced in the $K$-matrix formalism by Bowler {\it et al.} \cite{Bowler:1974th}. To justify this identification, we establish the connection between the formalism, introduced in the previous section, and the quark model.

\begin{enumerate}
	\item To make explicit the discussion in Ref.~\cite{Aitchison:1972ay}, we distinguish two types of interactions:

		\begin{itemize}
			\item The first type of interactions is described by Hamiltonian $H_0$, which describes the $q\overline{q}$ potential of the bound states of mesons, $\{a^0,b^0,\dots\}$. It generates the initial meson masses and wave functions which are used to calculate the matrix elements of meson decays in the quark model (see next item).
			\item The second type of interactions, described by Hamiltonian $H^\prime$, represents the interaction vertices connecting these bound states to the continuum of all possible states of two interacting mesons, $\{i,j,\dots\}$:

				\begin{equation}
					f_{a^0i}=\langle a^0|H^\prime|i\rangle
					\label{}
				\end{equation}
				We commonly call these vertex interactions ``couplings''. These couplings can be precisely calculated within the $^3P_0$ quark-pair-creation model. With adequate choice of phases of the wave functions of the bound states the couplings can be set to be real. 
		\end{itemize}
	\item No direct interaction is assumed between two mesons. Nevertheless, there is rescattering since a meson pair can annihilate into one bound state and then be created again from the decay of this bound state. This rescattering process can be iterated arbitrary number of times, what is equivalent to a resummation of meson loops between the initial and final vertices (see Fig.~\ref{fig:rescattering0}).
		\begin{figure}[h!]\centering
			\includegraphics[width=1\textwidth]{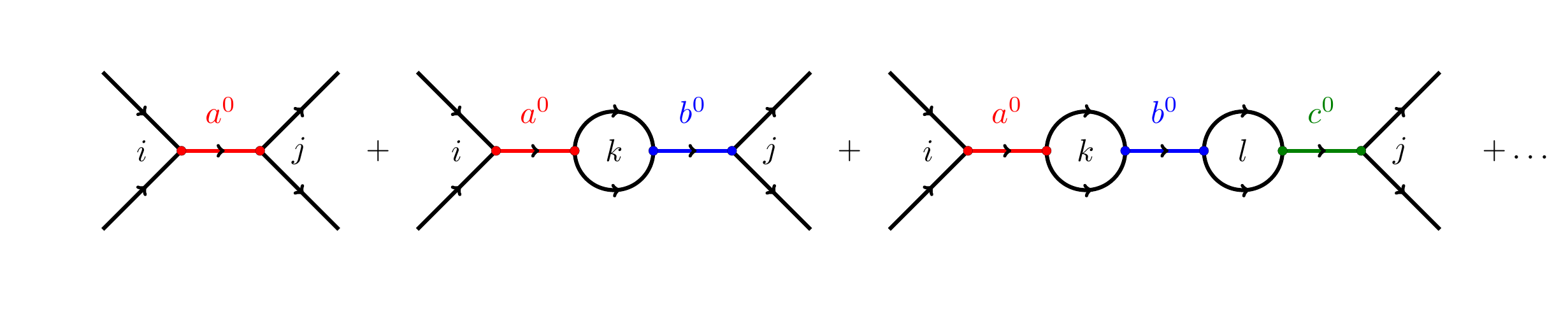}	
			\caption{Rescattering process.}
			\label{fig:rescattering0}
		\end{figure}

		All these possible processes can be resummed into a matrix propagator $\Pi_{a^0b^0}$ connecting two vertices. Let us call the scattering energy $m$. The couplings can be in principle energy dependent. i.e. be a function of $m$. This is the case with centrifugal barrier or damping factors, which are indeed present in our model. However, for simplicity of the presentation we assume them to be constant. Then, defining the ``bare'' scattering amplitude for the first diagram in Fig.~\ref{fig:rescattering0}
		
		\begin{equation}
			T_{ij}^{a^0(0)}=\frac{f_{ia^0}f_{a^0 j}}{m_{a^0}-m}
			\label{}
		\end{equation}
		and resumming all possible digrams with leads to the scattering amplitude

		\begin{equation}
			\begin{split}
				T_{ij} &= \sum_{a^0}T_{ij}^{a^0(0)}+\sum_{a^0,k,b^0} T_{ik}^{a^0(0)} I_k T_{kj}^{b^0(0)}+\sum_{a^0,k,b^0,l,c^0} T_{ik}^{a^0(0)} I_k T_{kl}^{b^0(0)} I_l T_{lj}^{c^0(0)}+\dots \\
				&= \sum_{a^0,b^0}f_{ia^0}\Pi_{a^0b^0}f_{b^0 j}
			\end{split}
			\label{eq:rescattering0}
		\end{equation}
		where the propagator defined as

		\begin{equation}
			\begin{split}
				\Pi_{a^0b^0}(m) &= \frac{\delta_{a^0b^0}}{m_{a^0}-m}+\frac{f_{a^0k}f_{kb^0}}{(m_{a^0}-m)(m_{b^0}-m)}I_k+\dots \\
				&=\left(M(m)-i\frac{\Gamma(m)}{2}-m\right)_{a^0b^0}^{-1}
			\end{split}
			\label{eq:Sprime1}
		\end{equation}
		with $I_k$ being the loop integral for the rescattering loop of the $k^\text{th}$ channel. In the case where the couplings $f$ depend actually on $m$, one should include the coupling factors relative to the internal lines of Fig. \ref{fig:rescattering0} in the loop integral. However, since  we do not attempt to calculate actually the loops, we see that there is  no need to introduce this complication.

		The mass matrix in \eqref{eq:Sprime1} is:
\begin{equation}
	\left(M-i\frac{\Gamma}{2}\right)_{a^0b^0}=m_{a^0} \delta_{a^0b^0}-\sum_i f_{ia^0}f_{b^0 i} I_i
\end{equation}		
It is in general non-diagonal. It contains
		\begin{itemize}
			\item the initial diagonal mass matrix $\text{diag}(m_{a^0},m_{b^0},\dots)$ of the bound states;
			\item the contribution of the loops for each possible channel, which can be non-diagonal since common two-body channels can couple to two different bound states. The loop integrals contain real and imaginary parts, which appear only when a two-body channel is open at the energy $m$.
		\end{itemize}

	\item Now the mass matrix must be diagonalized in two steps as explained in \cite{Aitchison:1972ay}. One first diagonalises the real part, $M$, then one passes to a diagonalization of the full new matrix, $M^\prime-i\Gamma^\prime/2$.

		\begin{enumerate}
			\item Diagonalization of the real part of the denominator of $\Pi_{a^0b^0}$ matrix, $M$, leads to the introduction of the new diagonal mass matrix% (see Eq.~\eqref{eq:diagonalization1})

				\begin{equation}
					M^\prime(m) = \text{diag}(m_{a^\prime},m_{b^\prime},\dots)
					\label{}
				\end{equation}

				This mass diagonalization implies a simultaneous rotation of the couplings $\{f_{a^0i}\}$, leading to the new couplings $\{f_{a^\prime i}\}$. One thus passes to the masses and couplings of the $K$-matrix, Eq.~\eqref{eq:Kmatrix0}.Of course, if there exists only one resonance which couples to the initial and final states, no rotation is needed. In this case all bare couplings $\{f_{a^0i}\}$ coincide with the ones of the $K$-matrix, $\{f_{a^\prime i}\}$. Thus, one can relate them with couplings calculated in the quark model. Otherwise, when there are two possible overlapping resonances, namely the two $K_1$'s, we have to make a rotation and introduce a mixing angle. We notice then that we have introduced an arbitrary rotation angle $\theta_{K_1}$ in our model computations which allows us to identify the set of the observed $K$-matrix couplings with the theoretical ones by the fit of data with our model predictions. This identification means that:
				\begin{itemize}
					\item the effect of the real part of the loops, i.e. $Re(I_k)$ in Eq.~\eqref{eq:rescattering0}, are taken into account in our model;
					\item mixing angle $\theta_{K_1}$ is not predicted by the model but is simply adjusted to data;
					\item introduction of the mixing angle $\theta_{K_1}$ can also take into account the uncalculated rotation of the pure spin states $K_{1A}$ and $K_{1B}$ into the eigenstates of Hamiltonian $H_0$ due to the spin-orbit forces~\cite{Godfrey:1986wj}.
				\end{itemize}
			\item The second step consists the diagonalization of the new mass matrix

				\begin{equation}
					\left(M^\prime-i\frac{\Gamma^\prime}{2}\right)_{a^\prime b^\prime}=m_{a^\prime}\delta_{a^\prime b^\prime}-i\sum_i\rho_{ii}(m)f_{a^\prime i}f_{ib^\prime}
					\label{eq:K-matrix_mass_matrix}
				\end{equation}
		\end{enumerate}
		This leads to the physical mass eigenstates and to the Breit-Wigner parametrization with energy-dependent width. This new rotation that accomplishes the last transformation into the physical states must have a complex and the angle of this rotation must have a complex phase. This would lead to the complex couplings of the mass eigenstates to set of continuum states. As we have already mentioned in the text, this rotation seems to be rather small.

	\item Let us now discuss the dependence of various variables on the energy $m$. In principle, all the masses and couplings, produced by two previous steps are dependent on $m$ because of the loop effects (the bare couplings themselves may depend on $m$, as is the case of our quark model, when one calculates the decay momenta, then the widths at the mass $m$. This then modifies the expression of the loop integral). This also implies that the real mixing angle $\theta_{K_1}$ is also energy dependent in principle. However, as regards the mass matrix, its real and imaginary parts have rather different behaviour depending on $m$. In first approximation, the real part of the mass matrix, which includes the sum of the large number of loops, varies slowly with $m$ and can be considered as constants on a limited range of energy. This is what was done in the analysis of Daum {\it et al.}. On the contrary, the imaginary part, which corresponds to the partial widths of the opened channels, is a rapidly varying function near the threshold.

		One can go beyond the approximation of the real part of the mass matrix by taking into account that there is some variation near the threshold. This is obtained by analytic continuation of the phase space through the threshold. This effect is consistently included in the prescription of Nauenberg and Pais of the complex phase space, although we differ on other assumptions they made. This corresponds to having imaginary part of the widths generating a $m$-dependent mass shift. For instance, for the $K$-matrix width one have

		\begin{equation}
			\Gamma_{a^\prime b^\prime}(m)=\sum_i\rho_{ii}(m)f_{a^\prime i}f_{ib^\prime}
			\label{}
		\end{equation}
		where the phase-space factor $\rho_{ij}(m)$ can be complex in general.
		
		Finally, one obtains for the physical states that the physical masses of $K_1(1270/1400)$ are varying slowly as functions of $m$ while the physical widths are rapidly changing functions; moreover the mass of $K_1(1270)$ has a more rapid variation around the peak due to the closeness of the $K_1$-mass to the $K\rho$ threshold (see  Fig.~\ref{fig:Mass_Width_diag}).
\end{enumerate}

In summary, one should identify the $K$-matrix couplings with the ones predicted in the $^3P_0$ model, with the real mixing effect included to define the initial states in this model. To establish the quantitative relation between the definitions in these two formalisms, we identify more exactly the \underline{\it partial widths}, $\Gamma_{K_1i}^{QPCM}(M_{peak})$:

\begin{equation}
\Gamma_{K_1i}^{QPCM}(M_{peak})=\Gamma_{K_1i}=2f_{K_1i}^2 Re[\rho_{ii}(M_{peak})]
	\label{eq:GammaK1_partial}
\end{equation}
where $\Gamma_{K_1i}^{QPCM}(M_{peak})$ is the $^3P_0$ model partial width, Eq. \eqref{eq:partial_width_QPCM} ; $\Gamma_{K_1i}$,  $\rho_{ij}$ and $f_{K_1i}$ are respectively the partial width. the phase space (we use the real part of the phase space since $\rho_{ij}$ is defined as a complex quantity as will be explained later) and the $K$-matrix couplings in the formalism of Daum {\it et al.} Note that these are not exactly the common partial widths related to the Breit-Wigner analysis: the latter would be obtained by applying the complex rotation (see Eq.~\eqref{eq:K-matrix_mass_matrix}).

Now, Eq.~\eqref{eq:partial_width_QPCM} is valid only for the narrow isobar. If we have to take into account the effect of the finite width of the isobar, we have to integrate the quasi-two-body phase space over the Breit-Wigner of the isobar. One has to underline, that in this approach we do not have to integrate over the Breit-Wigner of the $K_1$-resonance unlike what is done, for instance, in Ref.~\cite{Kokoski:1985is}. We indeed calculate the width at the peak. On the contrary, if we would like to compare with the results of the Belle collaboration analysis \cite{Belle:2010if}, this approach must be changed and we would have to integrate over the whole three-body phase space of $B \to K\pi\pi \psi$ to obtain the {\it branching ratios}. But, even in that case, it does not have sense, in our opinion, to integrate the decay widths themselves over the Breit-Wigner of the $K_1$.

\subsection{Observed problems in the experimental $K$-matrix analysis}\label{sec:problems}

As announced, we found several problems in using the experimental analysis:

\begin{itemize}
	\item {\bf Absence of the $K\pi\pi$ non-resonant contribution in the $K$-matrix}.

		We note that the $K$-matrix of Daum {\it et al.} is composed only of two resonance poles. There is no non-resonant contribution which is usually parametrized as polynomial in terms of $m$ in the $K$-matrix parametrization. This implies the strong assumption that the quasi-two-body scattering of vector-scalar mesons ($K^*\pi$ and $K\rho$) passes only through the $K_1$ resonant intermediate states.

	\item {\bf $D$-wave amplitudes issue}.
		
		The results of the ACCMOR analysis show that the $D$-wave in $K_1(1270)\to K^*\pi$ depends strongly on the production transfer $t$ in the $Kp\to K\pi\pi p$ reaction. This fact may escape the attention of PDG reader, because it averages between two sets of data (low $t$, high $t$). As for the $D$-wave amplitude in the $K\rho$ channel, there is no information; only branching ratios are quoted in the paper but not the $K$-matrix couplings and their phases which are crucial for our study.
	\newpage
	\item {\bf The problem of definition of the total width with threshold effect}
		
		When the mass of the resonance at the peak is close to a decay threshold, different definitions of the resonance width are no longer equivalent. Such possible definitions are the width at the peak $\Gamma(M_{peak})$, the width at the $S$-matrix pole, and finally the full width if measured at one-half the maximum height (FWHH) of the Breit-Wigner distribution defined as
		
		\begin{equation}
			\Gamma_{K_1}^{FWHH}\equiv m_2-m_1,
			\label{}
		\end{equation}
		where $m_1$ and $m_2$ are defined as two solutions in $m$ of the equation

		\begin{equation}
			\frac{f_{a^\prime(b^\prime)1}^2\rho_{11}(m)}{m_{a^\prime(b^\prime)}-m-i\Gamma_{a^\prime(b^\prime)}(m)}=\frac{1}{2}\frac{f_{a^\prime(b^\prime)1}^2\rho_{11}(M_{peak})}{m_{a^\prime(b^\prime)}-M_{peak}-i\Gamma_{a^\prime(b^\prime)}(M_{peak})}
			\label{}
		\end{equation}
		using the $K^*\pi$ channel (labelled as channel 1).
		
		The last two widths are found to be smaller than the first one. That is why the $K_1(1270)$ width, $\Gamma_{K_1(1270)}=(90\pm8)$~MeV/$c^2$~\cite{Daum:1981hb}, which is assumed to be defined as the full width if measured at one-half the maximum height of the Breit-Wigner distribution of $K_1$, is less by a factor 1.5-2 than the total width at the peak (see Table~\ref{tab:total_widths1}) which is computed using the $K$-matrix couplings and summing over all possible intermediate channels, i.e. 

		\begin{equation}
			\Gamma_{K_1}^{peak}\equiv 2\sum_i f_{K_1i}^2Re[\rho_{ii}(M_{peak})]
			\label{eq:Gamma_peak}
		\end{equation}
		We find, indeed, for the later to be of the order of 200~MeV/$c^2$ with the inclusion of the $\kappa\pi$ channel (see Table~\ref{tab:total_widths1}). As a consequence, one observes a large discrepancy between the two possible definitions of the partial width that can be extracted from data of the ACCMOR collaboration: the partial width, defined in a ``standard'' way as $\Gamma(K_1(1270)\to K\rho)=\Gamma_{K_1}\times \mathcal{B}(K_1(1270)\to K\rho)$, is less by a factor 2-3 compared to the partial width at the peak, defined from the $K$-matrix couplings (see Table~\ref{tab:partial_widths1}). The total width, defined by ACCMOR collaboration and tabulated in PDG, seems therefore to be misleading. It should not be used to compare with the quark model predictions. According to us, previous theoretical analyses (for instance, in Ref.~\cite{Blundell:1995au}) unduely used for experimental partial widths the product of branching ratios with this total width of $K_1(1270)$ quoted by PDG.

		\begin{table}[h!]
			\centering
			\begin{tabular}{|c|c|c|c|}
				\hline
				$K_1$ & $\Gamma_{K_1}^{ACCMOR}$, MeV/$c^2$ & $\Gamma_{K_1}^{peak}$, MeV/$c^2$ & $\Gamma_{K_1}^{FWHH}$, MeV/$c^2$ \\
				\hline
				$K_1(1270)$ & 90$\pm$8 & $\sim$190 & $\sim$80 \\
				\hline
				$K_1(1400)$ & 165$\pm$35 & $\sim$230 & $\sim$230  \\
				\hline
			\end{tabular}
			\caption{Experimental total decay widths, calculated using the fitted parameters from Ref.~\cite{Daum:1981hb}. In our opinion, only the widths calculated at the peak must be used to compute partial widths from the branching ratios. Note that the $D$-waves are not included in the $\Gamma_{K_1}^{peak}$ estimation.}
			\label{tab:total_widths1}
		\end{table}

		\begin{table}[h!]
			\centering
			\begin{tabular}{|l|c|c|}
				\hline
				Decay channel $i$ & \begin{tabular}{c}$\Gamma_{K_1i}=\mathcal{B}_{K_1i}\times\Gamma_{K_1}^{ACCMOR}$,\\MeV/$c^2$\end{tabular} & \begin{tabular}{c}$\Gamma_{K_1i}^{peak}=2f_{K_1i}^2 Re\rho_{ii}$,\\ MeV/$c^2$\end{tabular} \\
				\hline
				$K_1(1270)\to(K^*\pi)_S$ & 12$\pm$3 & 28$\pm$26 \\
				\hline
				$K_1(1270)\to(K\rho)_S$ & 41$\pm$10 & 122$\pm$28 \\
				\hline
				$K_1(1400)\to(K^*\pi)_S$ & 162$\pm$13 & 211$\pm$59 \\
				\hline
				$K_1(1400)\to(K\rho)_S$ & 2$\pm$2 & 20$\pm$25 \\
				\hline
			\end{tabular}
			\caption{Experimental partial decay widths, calculated using the fitted parameters from Ref.~\cite{Daum:1981hb}. As is is underlined before, only the values from the last column must be used.}
			\label{tab:partial_widths1}
		\end{table}

	\item {\bf The problem of the phase space, $\rho_{ij}$}

		In the expression of the $T$-matrix in the $K$-matrix formalism the phase space factor $\rho_{ij}$ is defined as

		\begin{equation}
			\rho_{ij}(m)=\frac{2k_i(m)}{m}\delta_{ij}
			\label{eq:rhoij1}
		\end{equation}
		
		\noindent Naively, $k_i$, is the break-up momentum for the two-body decay channel $i$. But, in fact, Bowler {\it et al.} used for $k_i$ a particular formulation, proposed by Nauenberg and Pais~\cite{Nauenberg:1962xx}, which tries to take  into account two important effects:
		\begin{itemize}
			\item The requirement of the analiticity of the amplitude. The simplest way to satisfy it is the analytic continuation of the phase space through the threshold:

				\begin{equation}
					\rho_{ij}(m)=\begin{cases}
						\frac{2k_i(m)}{m}\delta_{ij},~ ~ \quad\text{above threshold} \\
						\frac{2i|k_i(m)|}{m}\delta_{ij}, \quad\text{below threshold}
					\end{cases}
					\label{eq:rhoij_analytic_continuation}
				\end{equation}
				It is the basic idea of the so called ``Flatte model'' which  has been used to analyse the $a_0(980)$-decay into $\eta\pi$ and $K\overline{K}$ states, the resonance being very close to the $K\overline{K}$ decay threshold. Similarly, this effect is also present in the $K_1(1270)$-decays into $K\rho$ and $K^*\pi$ channels with the resonance being at the threshold of $K\rho$. This is not so relevant for the $K_1(1400)$-decays where the resonance is far above the thresholds.
			\item The effect of the isobar width. The peculiarity of the $K_1(1270)$ with respect to $a_0(980)$ case is that the two-body final state includes one unstable particle, the isobar $V$ ($V=\rho$ or $K^*$). In order to take into account the width of the isobar, it is logical to integrate the three-body phase space over the Breit-Wigner of the isobar:
				
				\begin{equation}
					k_i(m)=\int_{m_V^{min}}^\infty k_i(m,m_V)\frac{\Gamma_V/2\pi}{(M_V-m_V)^2+\frac{\Gamma_V^2}{4}}dm_V
					\label{eq:ki_integral_continuation}
				\end{equation}
				where $k_i(m,m_V)$ has its non-relativistic expression~\footnote{For the relativistic phase space Eq.~\eqref{eq:Sunitarity3} no longer defines a real $K$-matrix in the physical region. The reason  is that the relativistic momentum does not remain imaginary below the threshold due to an additional complex branch point $\propto\sqrt{m^2-(m_V-m_P)^2}$. Therefore Nauenberg and Pais in Ref.~\cite{Nauenberg:1962xx} restricted to non-relativistic case.}

				\begin{equation}
					k_i(m,m_V)=\sqrt{\frac{2m_Vm_P}{m_V+m_P}\left(m-m_V-m_P\right)}
					\label{eq:ki_NR}
				\end{equation}
				The infinite upper limit in Eq.~\eqref{eq:ki_integral_continuation} corresponds to the analytical continuation of $k_i$ below the threshold for $m_V>m-m_P$.
		\end{itemize}

		As an approximation to this integral, Nauenberg and Pais proposed to use the complex mass of the isobar, $M_V\to M_V-i\Gamma_V/2$, in the expression of the momentum $k_i(m,m_V)$. These two prescriptions lead to a complex phase space, defined as

		\begin{equation}
			\rho_{ij}(m) = \frac{2k_i(m)}{m}\delta_{ij}=\frac{2}{m}\sqrt{\frac{2M_Vm_P}{M_V+m_P}\left(m-M_V-m_P+i\frac{\Gamma_V}{2}\right)}_i\delta_{ij}
			\label{eq:rhoij_N-P}
		\end{equation}
		where $P$ ($P=K$ or $\pi$) is the final state pseudoscalar meson in the quasi-two-body decay. According to us this prescription of using a complex mass is not satisfactory for the $\rho$ and $K^*$, especially for $K_1(1270)\to K\rho$. Indeed, we found by direct integration of Eq.~\eqref{eq:ki_integral_continuation} that the results are quite different from the ones obtained using Eq.~\eqref{eq:rhoij_N-P}, especially the real part of $\rho_{ij}(m)$ which corresponds to the real phase space in the $K_1(1270)\to K\rho$ case (see Fig.~\ref{fig:rhoij}). The same observation was formulated by Frazer and Hendry~\cite{Frazer:1964zz} when the paper of Nauenberg and Pais was published. They pointed out that this approximation is valid only for the very narrow resonances. The failure of this approach is very worrying since it is basic for the whole analysis of Daum {\it et al.}. In order to cure this problem, we formulate the following assumption: as explained below, instead of identifying the $K$-matrix couplings themselves we assume that it is the product of the couplings squared and the phase space which is given in a correct way by the experiment, at least approximately.

		\begin{figure}[h]\centering
			\includegraphics[width=0.45\textwidth]{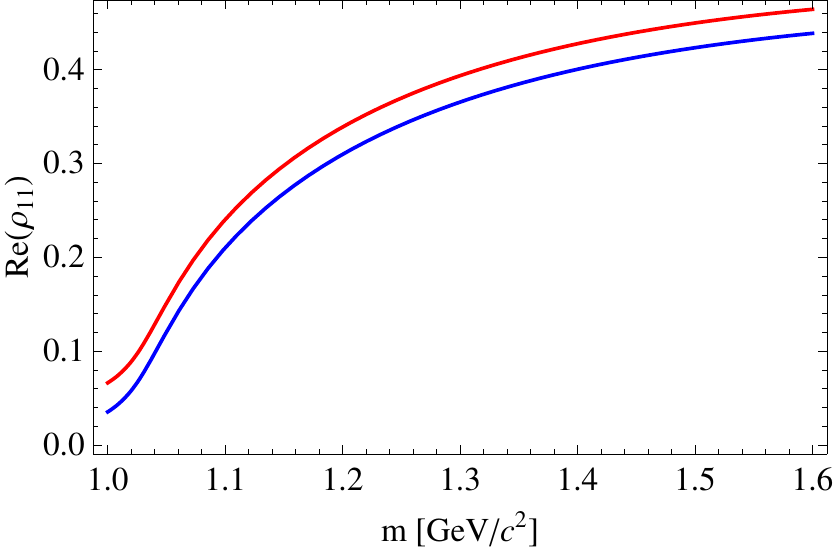}\hspace{5mm}
			\includegraphics[width=0.45\textwidth]{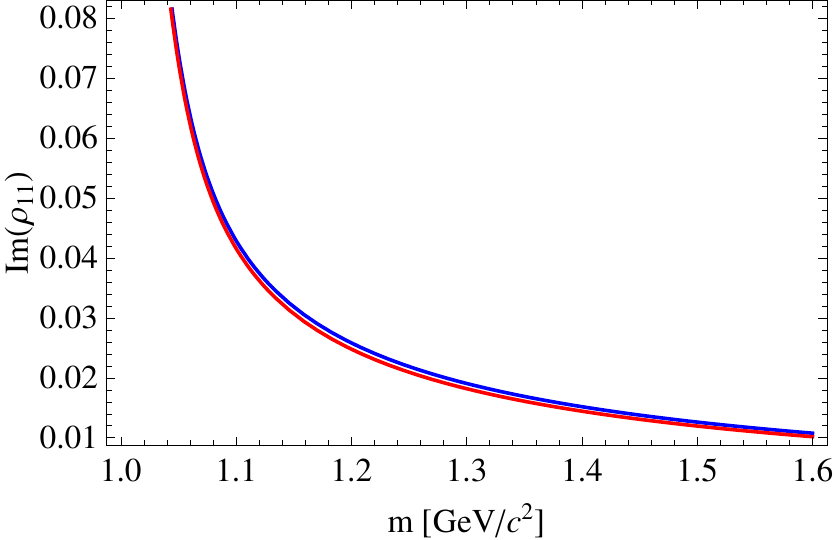}
			\vspace{5mm}
			\includegraphics[width=0.45\textwidth]{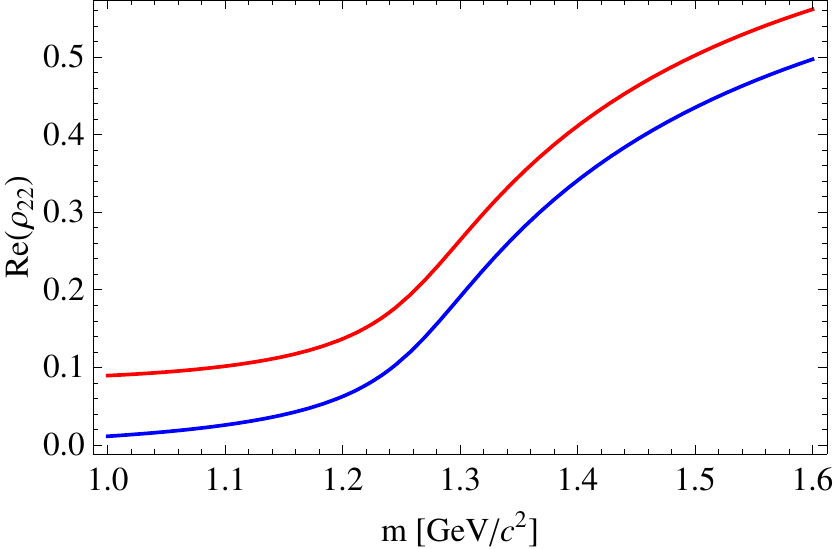}\hspace{5mm}
			\includegraphics[width=0.45\textwidth]{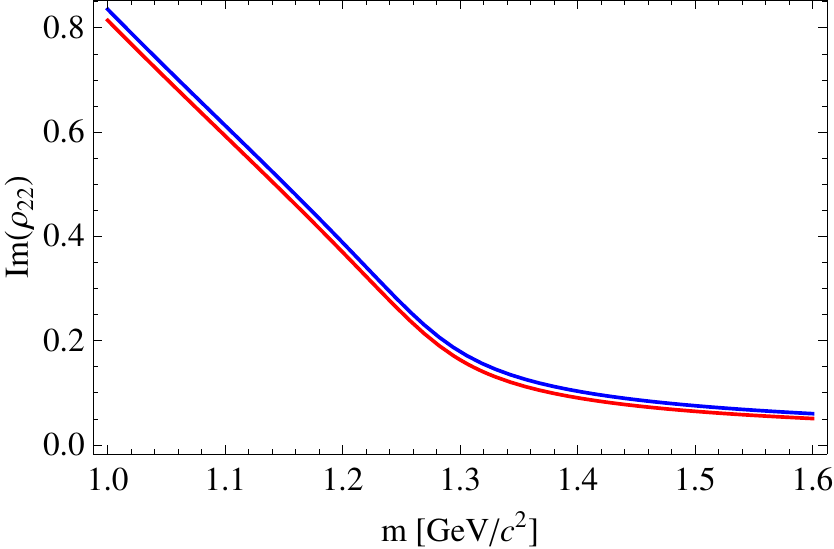}
			\caption{Dependence of the phase space factor $\rho_{ij}$ on the mass of the decaying resonance $m$ for the $K^*\pi$ (top) and $K\rho$ (bottom) channels. For comparison, $\rho_{ij}$ is calculated using the proper analytic continuation, Eq.~\eqref{eq:ki_integral_continuation}, (blue) and the approximation of Nauenberg and Pais, Eq.~\eqref{eq:rhoij_N-P}, (red). The difference between two approaches for $Re(\rho_{22})$ turns out to be significant for the $K\rho$ channel.}
			\label{fig:rhoij}
		\end{figure}
	\newpage
	\item {\bf The problem of the $P$- and $D$-waves}

		In addition, the prescription of Nauenberg and Pais has not been established for the $P$- and $D$-waves. We do not know what has been done exactly by Daum {\it et al.} to treat these waves. On the other hand, such waves are to be included in the analysis, especially the $\kappa\pi$ in the $P$-wave is very important. Since we are not able to redo the analysis by Daum {\it et al.} we use the couplings to $K_0^*(1430)\pi$ channel refitted by BABAR collaboration~\cite{Babar:2009ii}. They include a centrifugal barrier factor depending on the complex momentum which is defined by Eq.~\eqref{eq:rhoij_N-P}~\footnote{According to private communications.}. However, there is a new following problem here. The approximation of BABAR for the centrifugal barrier factor is not an approximation to the integral 

		\begin{equation}
			\int_{m_V^{min}}^\infty k_i(m,m_V)\left[\frac{k_i^2(m,m_V)\tilde{R}^2}{1+k_i^2(m,m_V)\tilde{R}^2}\right]\frac{\Gamma_V/2\pi}{(M_V-m_V)^2+\frac{\Gamma_V^2}{4}}dm_V
			\label{}
		\end{equation}
		which gives a positive real part while the approximation gives a negative one. This contradiction can be masked by the normalization of the centrifugal barrier factor at the peak. However, this is obviously not a satisfactory solution. 
	
	\item {\bf The diagonalization of the mass matrix and corresponding rotation of the $K$-matrix couplings into physical couplings}
		
		In several cases we have to deal not with the $K$-matrix couplings but with Breit-Wigner parametrization of the intermediate resonances. This is the case, for example, in our calculation of the $\mathcal{J}$-function. This is also the case of the Dalitz plot analyses such as the one of the Belle collaboration~\cite{Belle:2010if}. Then the relevant couplings are slightly different from those of the $K$-matrix. As stated before, they are obtained from the latter by a complex rotation. Indeed, to pass to the physical states we have to diagonalize the mass matrix of the states in the $K$-matrix formalism. This diagonalization can be performed by a complex orthogonal matrix. This rotation is complex because of the non-diagonal elements of the imaginary part of the mass matrix. The complex rotation angle (which depends on the energy) has both real and imaginary parts which are found to be of the order 10$^\circ$ (this result was obtained by explicit diagonalization of the mass matrix). As a consequence, this rotation affects the couplings: the rotation makes the couplings of the Breit-Wigner somewhat different from the ones of the real $K$-matrix. The magnitudes of the new couplings are different and phases appear. We found that the largest couplings (i.e. considering the dominant decay channels, $K_1(1270)\to K\rho$ and $K_1(1400)\to K^*\pi$) are slightly affected and acquire small phases. On the other hand for the smallest couplings ($K_1(1270)\to K^*\pi$ and $K_1(1400)\to K\rho$) the rotation effects are more important. In practical calculations of $\lambda_\gamma$ for the present moment we have neglected these effects so that we use directly the couplings obtained from the $^3P_0$ mode~\footnote{For more details, see the Appendix \ref{app:ACCMOR}}.
	
	\item {\bf Relative signs and ``offset'' phases}.

		It appears that the phases of the amplitudes, deduced from the \underline{\it experimental} $K$-matrix analysis are not exactly what is observed: this is a phenomenon of so-called ``offset'' phases. The $K\rho$ channel was found to have an additional unexplained phase of 30$^\circ$~\cite{Daum:1981hb} relative to the $(K^*\pi)_S$ which was set as a reference one. For the $\kappa\pi$ channel the discrepancy reaches 90$^\circ$.

		Another problem is that we are not able to establish the complete relation between the phase conventions of Daum {\it et al.} and quark model ones since the paper of ACCMOR collaboration is not detailed enough.
\end{itemize}

\section{Numerical results}\label{sec:results}

Let us summarize our final prescriptions we use for the calculation of the partial widths and for the further extraction of our theoretical model parameters from the experimental measurements. Our basic approach is to use {\it partial widths at the peak} on both, theoretical and experimental, sides. We abandon the idea of using the branching fractions and the total $K_1$-widths for the comparison with our predictions.

\begin{enumerate}
	\item For the theoretical prediction, in order to take into account the isobar width effects in our theoretical prediction of the partial widths $\Gamma_{K_1i}^{QPCM}$, the amplitudes \eqref{eq:K1SDamplitudes} squared are integrated over the invariant mass of the isobar:

		\begin{equation}
			\Gamma_{K_1i}^{QPCM}=8\pi^2\int_{m_V^{min}}^{M_{K_1}-m_P}\frac{E_VE_Pk_P}{M_{K_1}}|A_i(K_1\to VP)|^2\frac{\Gamma_V/2\pi}{(M_V-m_V)^2+\frac{\Gamma_V^2}{4}}dm_V
			\label{}
		\end{equation}
		Note that since we consider the widths at the peak there is no integration over the $K_1$ invariant mass unlike what is done in  several theoretical papers (e.g. see Ref.~\cite{Kokoski:1985is}). Moreover, one can notice that the integration over the mass of the isobar is one within the correct physical region restricted by the corresponding physical bound of the two-body decay (i.e. we use the real phase space).

	\item For the experimental input, we make the simple assumption that the partial widths, calculated from the $K$-matrix couplings at the peak according to Eq.~\eqref{eq:partial_width_Kmatrix}, are correct, although the complex phase space {\it \`a la} Nauenberg and Pais \eqref{eq:rhoij_N-P} might be not correct (i.e. what we measure by fitting data, is always the combination like $f_{a^\prime(b^\prime)i}^2\times\rho_{ij}(m)$ which are assumed to be extracted correctly). Therefore, we use the $K$-matrix couplings and the real part of the complex phase space \`a la Nauenberg and Pais in order to extract the experimental values of the partial widths

		\begin{equation}
			\Gamma_{K_1i}^{peak} = 2f_{a^\prime(b^\prime)i}^2Re[\rho_{ij}(M_{peak})]
			\label{eq:partial_width_Kmatrix}
		\end{equation}
	
	\item We calculate this partial width according to Eq.~\eqref{eq:partial_and_total_widths_K} also for the $P$ ($L=1$) and $D$-waves ($L=2$), assuming that the $K$-matrix couplings $f$ contain the barrier factors $B_i^L(m)$ that are properly normalized at the peak:
				
				\begin{equation}
					\begin{split}
						\left.f_{a(b)i}(m)\right|_{P,D-\text{waves}} &= f_{a(b)i}\frac{B_i^L(m)}{B_i^L(M_{peak})} \\
						B_i^L(m) &= \left[\frac{k_i^2(m)\tilde{R}^2}{1+k_i^2(m)\tilde{R}^2}\right]^{L/2}
					\end{split}
					\label{}
				\end{equation}
				where $\tilde{R}^2=25$~GeV$^{-2}$~\cite{Babar:2009ii}. This assumption seems to be correct since it leads to the calculated branching ratios  that are very close to the ones announced in the paper by Daum {\it et al.}. In any case we avoid as much as possible to rely on the experimental data on $K_1(1270)\to K\rho$ and the $D$-wave of $K_1(1270)\to K^*\pi$ and we trust our theoretical prediction.
\end{enumerate}

\subsection{Fit of parameters $\gamma$ and $\theta_{K_1}$} \label{sec:parameters}

In order to extract our phenomenological parameters, the quark-pair-creation constant $\gamma$ and $K_1$ mixing angle, we do a fit using the method of least squares. As an experimental input we use the partial widths (namely, $\Gamma_{K_1i}^{peak}$ from Table~\ref{tab:partial_widths1}) only of the following processes: $K_1(1270)\to(K^*\pi)_S$, $K_1(1400)\to(K^*\pi)_S$, $K_1(1400)\to(K\rho)_S$, which are assumed to be Gaussian distributed with mean $\Gamma_{K_1i}^{QPCM}(\gamma,\theta_{K_1})$ and known variance $\sigma_{\Gamma_{K_1i}^{peak}}$. The $D$-waves are not taken into account in our fit. Moreover, the dominant channel $K_1(1270)\to K\rho$ due to the dangerous threshold and phase space effects is avoided since the narrow width approximation can be incorrect for the decays near the threshold and here the width effects can play a significant role.

Then, the likelihood function is constructed as a sum of squares

\begin{equation}
	\chi^2(\gamma,\theta_{K_1})=-2\ln L(\gamma,\theta_{K_1})=\sum_{i=1}^3\frac{(\Gamma_{K_1i}^{peak}-\Gamma_{K_1i}^{QPCM}(\gamma,\theta_{K_1}))^2}{\sigma_{\Gamma_{K_1i}^{peak}}^2}
	\label{eq:chi2}
\end{equation}

In order to find the unknown parameter $\theta_{K_1}$ the function $\chi^2$ is minimized, or equivalently the likelihood function $L(\theta_{K_1})$ is maximized. The minimization of the $\chi^2$ gives the minimal value $\chi_\text{min}^2=0.61$ and the estimators $\hat\gamma=4.0$ and $\hat\theta_{K_1}=59^\circ$.

The covariance matrix for the estimators $\mathcal{V}_{ij}=\text{cov}[\hat\xi_i,\hat\xi_j]$ can be found from

\begin{equation}
	(\mathcal{V}^{-1})_{ij}=\left.\frac{1}{2}\frac{\partial^2\chi^2}{\partial\xi_i\partial\xi_j}\right|_{\xi=\hat\xi}
	\label{}
\end{equation}
Thus one obtains

\begin{equation}
	\text{cov}[\hat\gamma,\hat\theta_{K_1}]=
	\left(\begin{array}{cc}
		\sigma_\gamma^2 & C_{\gamma\theta_{K_1}} \\
		C_{\gamma\theta_{K_1}} & \sigma_{\theta_{K_1}}^2
	\end{array}\right)=
	\left(\begin{array}{cc}
		0.29 & 0.99 \\
		0.99 & 107.0
	\end{array}\right)
	\label{}
\end{equation}
where the diagonal elements give the variances $\sigma_{\hat\gamma}^2$ and $\sigma_{\hat\theta_{K_1}}^2$. Finally, one finds the fitted values of the quark-pair-creation constant and $K_1$ mixing angle:

\begin{equation}
	\gamma\simeq4.0\pm0.5, \quad\quad \theta_{K_1}\simeq(59\pm10)^\circ
	\label{eq:fitted_gamma_thetaK1}
\end{equation}

Taking for granted that our theory is correct, one is now interested in the quality of the agreement  between data and various realizations of the theory, determined by the set of parameters, namely $\{\gamma,\theta_{K_1}\}$. For metrological purposes one should attempt to estimate as best as possible the complete set of parameters $\{\gamma,\theta_{K_1}\}$. In this case we use the offset-corrected $\chi^2$ \cite{Charles:2004jd}:

\begin{equation}
	\Delta\chi^2(\gamma,\theta_{K_1})=\chi^2(\gamma,\theta_{K_1})-\chi_\text{min}^2
	\label{}
\end{equation}
where $\chi_\text{min}^2$ is the absolute minimum value of the $\chi^2$ function of Eq.~\eqref{eq:chi2} which is obtained when letting our model parameters free to vary. The minimum value of $\Delta\chi^2$ is zero by construction. Here one has to notice, that this absolute minimum does not correspond to a unique choice of the model parameters. This is due to the fact that the theoretical predictions used in the analysis are affected by important theoretical systematical errors. Since these systematics are restricted in the allowed regions there is always a multi-dimensional degeneracy for any value of $\chi^2$. However, since in our analysis there are only two model parameters, our predictions for $\{\gamma,\theta_{K_1}\}$ are not affected by any other theoretical predictions.

\begin{figure}[h!]\centering
	\includegraphics[width=0.45\textwidth]{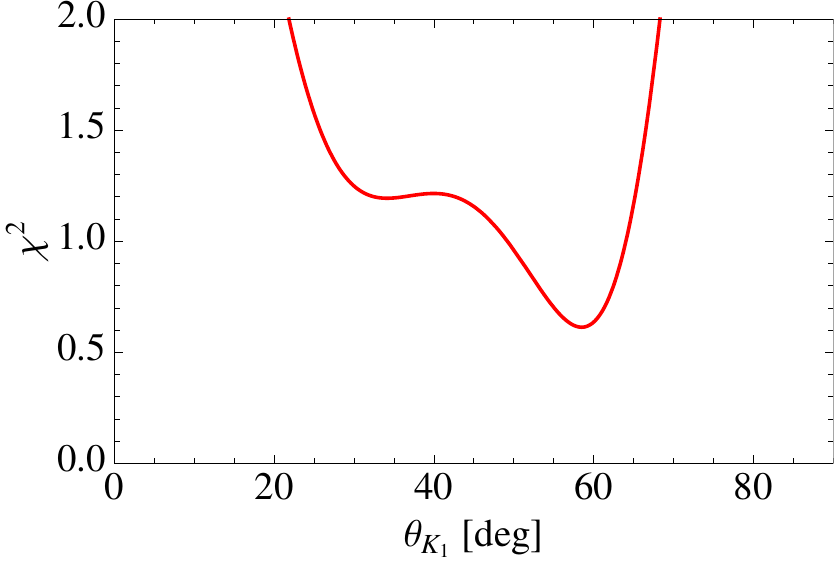}\hspace{5mm}
	\includegraphics[width=0.45\textwidth]{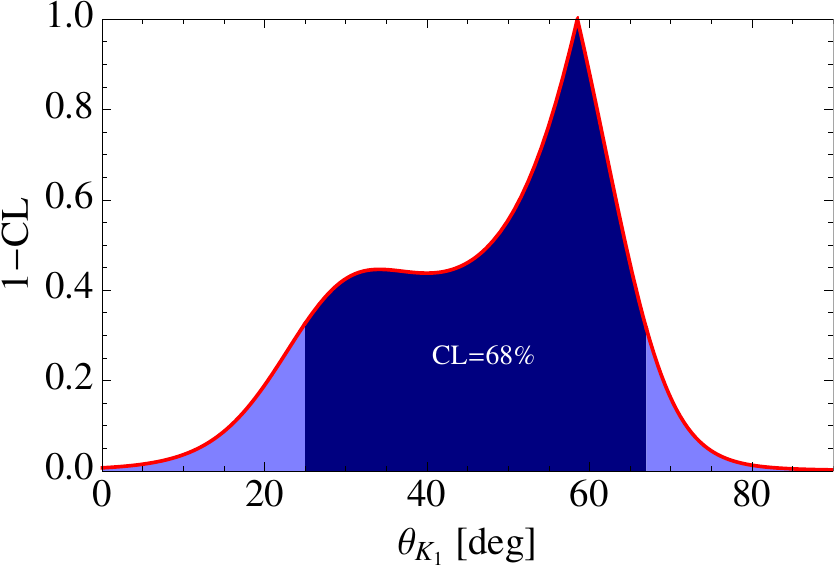}
	\vspace{5mm}
	\includegraphics[width=0.45\textwidth]{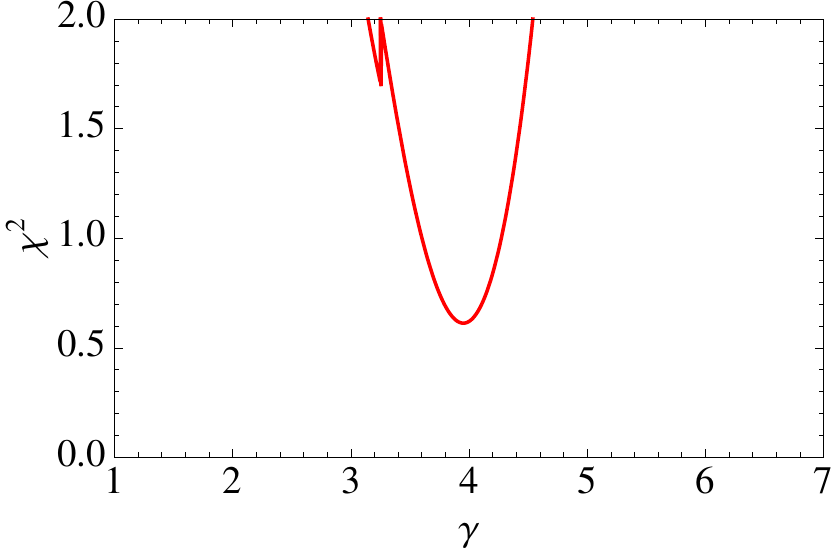}\hspace{5mm}
	\includegraphics[width=0.45\textwidth]{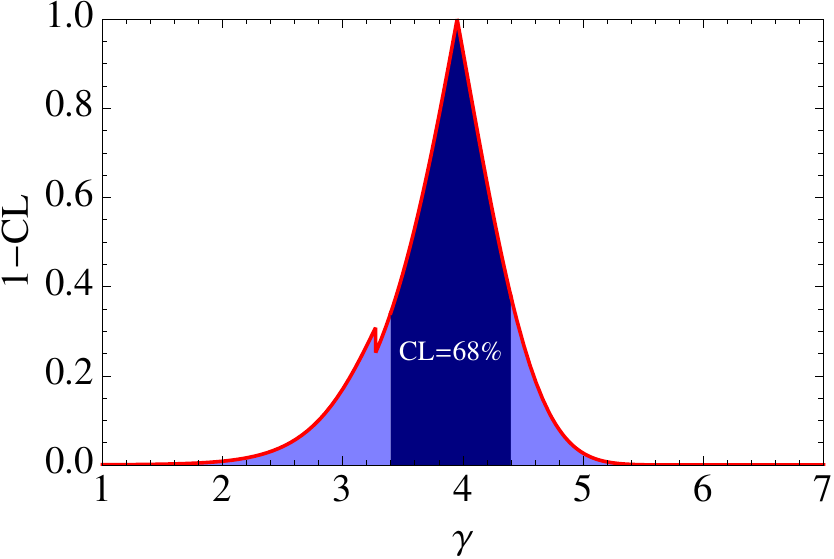}
%	\vspace{5mm}
%	\includegraphics[width=0.45\textwidth]{gamma-thetaK1_1sigma_68.pdf}
	\caption{$\chi^2$ distributions for the fitted parameters, $K_1$ mixing angle $\theta_{K_1}$ and quark-pair-creation constant $\gamma$ (left), with the confidence level intervals that determine how frequently the observed interval contains the parameters (right).}
	\label{fig:gamma_thetaK1_CL}
\end{figure}

A necessary condition is that the confidence level (CL) constructed from $\Delta\chi^2(\gamma,\theta_{K_1})$ provides correct coverage is that the CL interval~\footnote{In statistics, a confidence level interval is a particular kind of interval estimate of a fitted parameter and is used to indicate the reliability of an estimate. It is an observed interval (i.e. it is calculated from the observations), in principle different from sample to sample, that frequently includes the parameter of interest, if the experiment is repeated. How frequently the observed interval contains the parameter is determined by the confidence level.} for $\{\gamma,\theta_{K_1}\}$ covers the true parameter value with a frequency 1-CL if the measurements were repeated many times. The corresponding CL intervals for the confidence level of CL=68\% are shown in Fig.~\ref{fig:gamma_thetaK1_CL}.

%\begin{figure}[h!]\centering
%	\includegraphics[width=0.4\textwidth]{gamma-thetaK1_regions.pdf}
%	\caption{QPCM constraints for the quark-pair-creation constant $\gamma$ and the $K_1$ mixing angle $\theta_{K_1}$ obtained from the fitted partial decay widths at the peak, calculated using the $K$-matrix couplings (Table~\ref{tab:partial_widths1}). The cross indicates the optimal values of $\gamma$ and $\theta_{K_1}$ extracted from the fit.}
%	\label{fig:gamma-thetaK1}
%\end{figure}

\subsection{Model predictions for partial widths}

\begin{figure}[h!]\centering
	\includegraphics[width=0.45\textwidth]{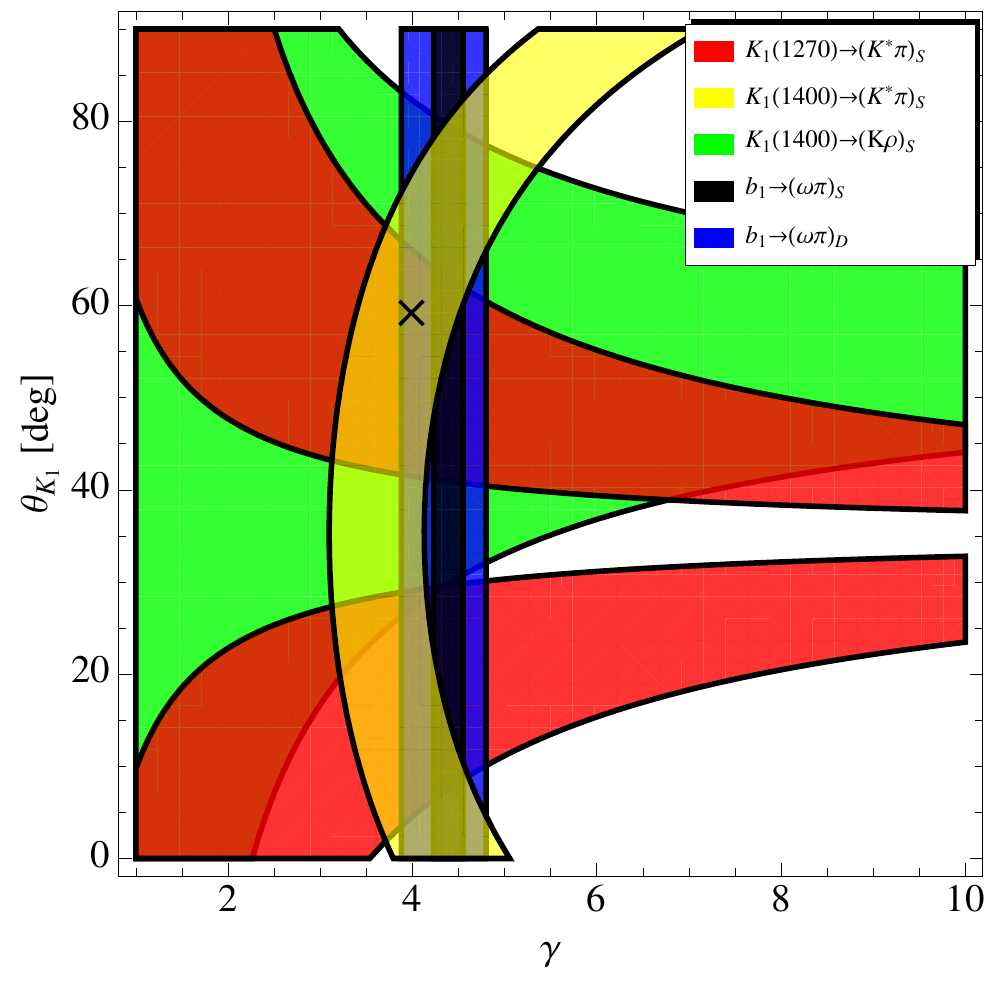}
	\caption{QPCM constraints for the quark-pair-creation constant $\gamma$ and the $K_1$ mixing angle $\theta_{K_1}$ obtained from the fitted partial decay widths at the peak, calculated using the $K$-matrix couplings (Table~\ref{tab:partial_widths1}). The cross indicates the optimal values of $\gamma$ and $\theta_{K_1}$ extracted from the fit.}
	\label{fig:gamma-thetaK1}
\end{figure}

Now, we can make systematic predictions for various processes. First, it is very useful to check our result for the quark-pair-creation constant $\gamma$ prediction with the much better studied $b_1\to(\omega\pi)_S$ and $b_1\to(\omega\pi)_D$ decays~\footnote{One has to point out that the branching ratio of $b_1\to\omega\pi$ has not been measured precisely. However, the $\omega\pi$ is considered to be the dominant decay mode~\cite{PDG}, so that we assume $\mathcal{B}(b_1\to\omega\pi)\simeq100\%$.} which depend only on $\gamma$. One can see from Fig.~\ref{fig:gamma-thetaK1} that our estimation for $\gamma$, determined from the $K_1$-decays \eqref{eq:fitted_gamma_thetaK1}, is in a good agreement with the one extracted from the $b_1\to\omega\pi$ decay. Moreover, the extracted $D/S$ ratio of the partial amplitudes is very well predicted and coincides with the measured value including the sign:
\begin{equation}
 (A_D/A_S)_{QPCM}\simeq0.28
\end{equation} 
while the experiment~\cite{PDG} gives:
\begin{equation}
 (A_D/A_S)_\text{exp}=0.277\pm0.027
\end{equation}
Note that the Belle collaboration omits the $D$-waves in the $B\to J/\psi K_1$ analysis. This could be of consequence, since the Dalitz plot should be appreciably different according to our calculation (see our discussion in the end of subsubsection \ref{BellePsi})

To summarize, we give in Table~\ref{tab:partial_widths_QPCM} our predictions for the $S$-wave partial widths of the strong interaction decays of the $K_1$-mesons, using the fitted values of $\gamma$ and $\theta_{K_1}$. One can see that the agreement is satisfactory except for the $K_1(1270)\to K\rho$ channel. This is not unexpected in view of the particular difficulties of the experimental treatment in this decay as explained in the previous section (recall especially that the drawback of using the phase space formula of Nauenberg and Pais is crucial in this case) .

\begin{table}[h!]\centering
	\begin{tabular}{|l|c|c|}
		\hline
		Decay channel $i$ & $\Gamma_{K_1i}^{QPCM}$, MeV/$c^2$ & $\Gamma_{K_1i}^{peak}$, MeV/$c^2$ \\
		\hline
		$K_1(1270)\to(K^*\pi)_S$ & 31 & 28$\pm$26 \\
		\hline
		$K_1(1270)\to(K\rho)_S$ & 61 & 122$\pm$28 \\
		\hline
		$K_1(1400)\to(K^*\pi)_S$ & 209 & 211$\pm$59 \\
		\hline
		$K_1(1400)\to(K\rho)_S$ & 1 & 20$\pm$25 \\
		\hline
	\end{tabular}
	\caption{Theoretical predictions for the partial decay widths, calculated using the fitted parameters $\gamma=4.0$ and $\theta_{K_1}=59^\circ$ and compared to the experimental partial values of widths at the peak (see Table~\ref{tab:partial_widths1}).}
	\label{tab:partial_widths_QPCM}
\end{table}

As for the $D$-waves in the $K_1$-decays, our impression is that they are poorly determined experimentally. Our prediction ($\Gamma(K_1(1270)\to(K^*\pi)_D)\simeq3$~MeV/$c^2$) lies below the experimental numbers: the couplings for the $D$-waves are not given in the paper by Daum {\it et al.}. Tentatively they were re-fitted by the BABAR collaboration~\cite{Babar:2009ii} from which we deduce the partial width $\Gamma(K_1(1270)\to(K^*\pi)_D)=(34\pm3)$~MeV/$c^2$. Here one has to notice that the errors of the re-fitted parameters are surprisingly small, as the ones obtained by Daum {\it et al.}.

\subsection{Prediction of signs of decay amplitudes and the ``offset'' phase issue \label{sec:offset_phases}}

Let us recall that, at least  for the determination of the photon polarization parameter $\lambda_\gamma$ as described in our paper \cite{Kou:2010kn}, our goal is to calculate the $\mathcal{J}$-function~\eqref{eq:J-function} which describes the full three-body $K_1\to K\pi\pi$ decay. As explained, we need in fact the expression $Im[\vec{n}\cdot(\vec{\mathcal{J}}\times\vec{\mathcal{J}}^*)]$  which depends crucially on the relative phases of the $V\to PP$ couplings and the $K_1\to VP$ form factors (see Eqs.~(22)-(27) in Ref.~\cite{Kou:2010kn} for the definition). These quantities are directly related to the two-body decay amplitudes, calculated by using the quark model. The phases of these amplitudes do not make sense by themselves but only in the product of two amplitudes of the subsequent processes which describe the final three-body decay $K_1\to K\pi\pi$. Then, the relative signs are observable quantitities, that can also be determined from any careful experimental study of.the $K_1$ decays. We define the relative phases for two $K_1\to K\pi\pi$ amplitudes of various partial waves via different intermediate isobar states (i.e. $(K^*\pi)_S$, $(K^*\pi)_D$, $(K\rho)_S$). Standardly, the reference partial wave is chosen to be the $S$-wave of $K^*\pi$. For instance, the relative phase of the $K_1\to K\rho\to K\pi\pi$ channel is defined as:
\begin{equation}
	\delta_\rho\equiv\text{arg}\left[\frac{A_S(K_1\to K\rho)\times A_P(\rho\to\pi\pi)}{A_S(K_1\to K^*\pi)\times A_P(K^*\to K\pi)}\right]
	\label{delta}
\end{equation}.

\noindent One has to note, that the total relative phase which is contained in the $\mathcal{J}$-function contains of course complex the phase of the denominator of Breit-Wigner of the isobar. For the conventions necessary to define $\delta_\rho$ we refer to Appendix. 

$\delta_\rho$ is independent of the conventional phase factors of the meson states (e.g. meson wave functions~\footnote{In the QPCM, $\delta_\rho$ can be calculated from
\begin{equation}
	\delta_\rho\propto\text{arg}\left[\frac{\psi^{(K_1)}\psi^{(K)*}\psi^{(\rho)*}\times \psi^{(\rho)}\psi^{(\pi)*}\psi^{(\pi)*}}{\psi^{(K_1)}\psi^{(K^*)*}\psi^{(\pi)*}\times \psi^{(K^*)}\psi^{(K)*}\psi^{(\pi)*}}\right]=\text{arg}\left[\frac{\psi^{(K_1)}\psi^{(K)*}\psi^{(\pi)*}\psi^{(\pi)*}}{\psi^{(K_1)}\psi^{(K)*}\psi^{(\pi)*}\psi^{(\pi)*}}\right]=1
	\label{}
\end{equation}
what implies that the relative phase of the {\it total} amplitudes is real (i.e. $\delta_\rho=0$ or $\pi$) and does not depend on the separate complex phases of the meson wave functions.}). In the $^3P_0$ model each decay amplitude is real with suitable conventions of the wave functions and by factorization of spherical harmonics. Then in the quark model $\delta_\rho$ is {\it real}. This is due to specific properties of the transition operator.

\subsubsection{Sign of the $D/S$ ratio}

The simplest prediction is the one concerning the $D/S$ ratio in the $b_1\to\omega\pi$ and $a_1\to\rho\pi$ decays. Indeed, this sign depends only on the well known standard conventions. It is then striking that all the signs are correctly predicted by the model. In the case of $b_1$ and $a_1$ these signs are well measured and given in PDG. For the $K_1\to K^*\pi$ channel the signs are not given by Daum {\it et al.} in~\cite{Daum:1981hb}. However, we can read the relative phase for $K_1(1270)$ from Fig.~(13) in Ref.~\cite{Daum:1981hb} which is positive ($f_{b5}/f_{b1}>0$), while for $K_1(1400)$ we have to rely on the analysis of BABAR because it is not possible to fix it from the figure since the $D$-wave is too weak overwhelmed compared to the $D$-wave of $K_1(1270)$ ($f_{a5}/f_{a1}<0$).

\begin{figure}[h]\centering
	\includegraphics[width=0.5\textwidth]{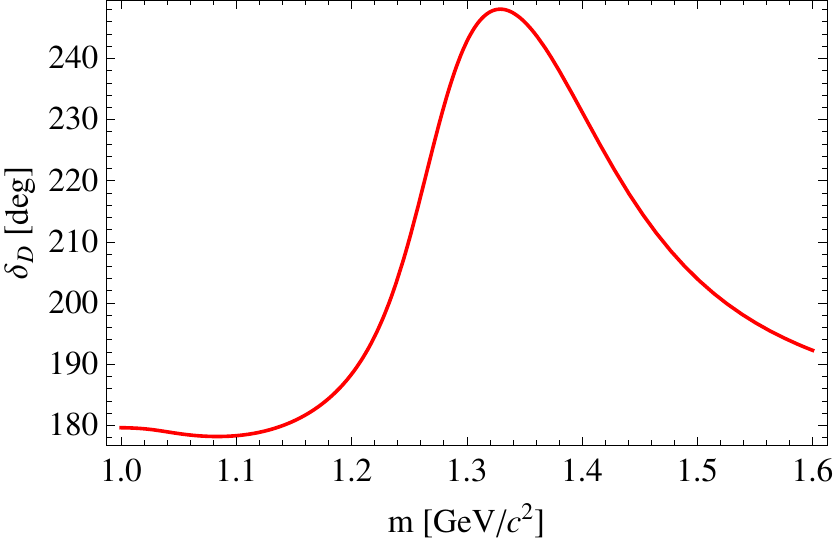}
	\caption{The $D$-wave phase relative to the $S$-wave of $K^*\pi$, calculated using the $K$-matrix couplings. One can see a bump at $M_{K\pi\pi}\sim(1.3-1.4)$~GeV/$c^2$.}
	\label{fig:D-wave_phase}
\end{figure}

In the paper of Gronau {\it et al.}~\cite{Gronau:2002rz,Gronau:2001ng} the $D/S$ phase for $K^*\pi$ is given as $\delta_{D/S}=(260\pm20)^\circ$. We believe that the authors were misled by incorrect interpretation of Fig.~(13) (bottom-right) in \cite{Daum:1981hb}: the plotted phase indeed peaks at 260$^\circ$ at $M_{K\pi\pi}\approx1.4$~GeV/$c^2$ . But this is not the phase we are looking for since it contains the phase from the Breit-Wigner of $K_1(1270)$ which is dominating over the $K_1(1400)$ contribution and gives an additional phase of approximately 90$^\circ$.  Hence, the phase we are interested in must be read as $\delta_{D/S}\approx(260-90)^\circ\sim180^\circ$. We must stress the following subtle point: the plotted phase is the difference of the phases of the $D$-wave strongly dominated by $K_1(1270)$ and the one of the $S$-wave which includes large contributions of both resonances. As a consequence, paradoxically, there appears a bump in the $D$-wave phase diagram, peaked at $M_{K\pi\pi}\sim(1.3-1.4)$~GeV/$c^2$ which is essentially determined by the tail of the Breit-Wigner of $K_1(1270)$. We checked this conclusion by explicit calculation of the amplitudes using the $K$-matrix couplings (see Fig.~\ref{fig:D-wave_phase}).

\subsubsection{Relative sign of the $K\rho/K^*\pi$ couplings}

We study the real phase (i.e. the relative sign) of the $K_1(1270)\to K^*\pi$ and $K_1(1270)\to K\rho$ amplitudes, which plays important role in the $\lambda_\gamma$ determination using the $\omega$-method (due to the strong dependence on the phase of the interference term $Im[\vec{n}\cdot(\vec{\mathcal{J}}\times\vec{\mathcal{J}}^*)]$). Indeed, the odd moments of $\omega$ change their sign if one changes the relative sign between the $K_1^+\to K^+\rho^0\to K^+\pi^-\pi^+$ and $K_1^+\to K^{0*}\pi^+\to K^+\pi^-\pi^+$ amplitudes. One has to notice that in this case this phase can be hardly extracted from the $K$-matrix analysis by Daum {\it et al.} due to some unknown conventions (in particular, the order of particles what is significant for the determination of the couplings signs). We then rely on the recent analysis by the Belle collaboration of the $B\to J/\psi(\psi^\prime)K\pi\pi$ decay which gives more explicit explanation of the conventions.

\vspace{5mm}
Here we summarise what is new in the Belle $B\to J/\psi(\psi^\prime)K\pi\pi$ paper~\cite{Belle:2010if}. First we will list up the general conclusions of this paper and then, discuss some details of the Dalitz plot shown in this paper, which provides important information to our work.

\subsubsection{General conclusions of the study of $B\to J/\psi K\pi\pi$ by the Belle collaboration}\label{BellePsi}

This paper, in principle, focuses on the measurement of the branching ratios of $B^+\to J/\psi K^+\pi^+\pi^-$ and $B^+\to \psi' K^+\pi^+\pi^-$. Since the $K\pi\pi$ final state comes from various resonances, $K_{\rm res}$, this analysis provides information of the $K_{\rm res}\to K\pi\pi$ strong decays. Since the $K_{\rm res}=K_1(1270)$ turned out to be a prominent component (for both $J/\psi$ and $\psi^{\prime}$), some detailed study of $K_1(1270)\to K\pi\pi$ has been done:
\begin{itemize}
	\item The Dalitz plot for the three-body decays is shown. We discuss more details on this later.

	\item The intermediated two-body decay branching ratios have been re-determined (see Table~\ref{tab:BR_Belle}). The branching ratios for the dominant decay modes, $K_1(1270)\to K\rho$ and $K_1(1270)\to K^*\pi$, are found to be slightly different from the previous measurements (PDG), although they are still in accordance within several standard deviations. On the other hand, the $K_1(1270)\to K_0^*(1430)\pi$ channel, which was supposed to have a large branching fraction ($\mathcal{B}(K_1(1270)\to K_0^*(1430)\pi)=(28\pm4)\%$) according to the previous measurements~\cite{Daum:1981hb,PDG}, was found to have a significantly smaller contribution of the order of $2\%$ (see Table~\ref{tab:BR_Belle}).
	\item In addition, by floating the mass and width of the $K_1(1270)$ in an additional fit of the $B^+\to J/\psi K^+\pi^+\pi^-$ data, a smaller mass of $(1248.1\pm3.3(\text{stat})\pm1.4(\text{syst}))$~MeV/$c^2$ and larger width $(119.5\pm5.2(\text{stat})\pm6.7(\text{syst}))$~MeV/$c^2$ were measured for the $K_1(1270)$. Of course, there is a correlation between the fact that the ``scalar+$\pi$'' component becomes much smaller and the fact  that the $K^*\pi$ and $K\rho$ contributions become larger (see Table~\ref{tab:BR_Belle}).
\end{itemize}
			
Here we want to draw attention of the reader to the conceptual difficulties raised by the definition of the $K_1(1270)$-width. In the Fit~1 the $K_1$ width is the one given by PDG while in the Fit~2 the width was treated as a free parameter. Due to the threshold effect one should not expect that the width measured by the Belle collaboration from the Breit-Wigner denominator at the peak should coincide with the one defined by PDG, although it should be much larger. One observes that the floated width is larger than the PDG value but it is still much smaller than 200~MeV/$c^2$ as we would expect from the calculation using the $K$-matrix formalism (see Table~\ref{tab:total_widths1}).

One has to point out  that the $D$-waves are not taken into account in the master formula of Belle. On the other hand, we found from the theoretical study that the $D$-wave of $K^*\pi$ can have a small but non-negligible effect. In principle, there are two bumps due the presence of the $D$-wave, but it is found that the one located in the intersection region of the $M_{K\pi}\sim M_{K^*}$ and $M_{\pi\pi}\sim M_\rho$ on the Dalitz plot is masked by the dominating peak of $\rho$. Using a Monte-Carlo simulation, we observed a second small but non-negligible bump at low $M_{\pi\pi}$ (see Fig.~\ref{fig:Dalitz_pipi-Kpi} in the center).

\begin{table}\centering
	\begin{tabular}{|c|c|c|c|}
		\hline
		Decay mode & PDG ($\%$) & Fit 1 ($\%$) & Fit 2 ($\%$) \\
		\hline
		$K\rho$ & $42\pm6$ & $57.3\pm3.5$ & $58.4\pm4.3$ \\
		\hline
		$K^*\pi$ & $16\pm5$ & $26.0\pm2.1$ & $17.1\pm2.3$ \\
		\hline
		$K_0^*(1430)\pi$ & $28\pm4$ & $1.90\pm0.66$ & $2.01\pm0.64$ \\
		\hline
	\end{tabular}
	\caption{The fitted branching ratios of the $K_1$-decays measured by the Belle collaboration in the analysis of $B\to J/\psi K\pi\pi$ decay \cite{Belle:2010if}. }
	\label{tab:BR_Belle}
\end{table}

\subsubsection{Dalitz analysis}

\begin{figure}[h]\centering
		\includegraphics[width=0.3\textwidth]{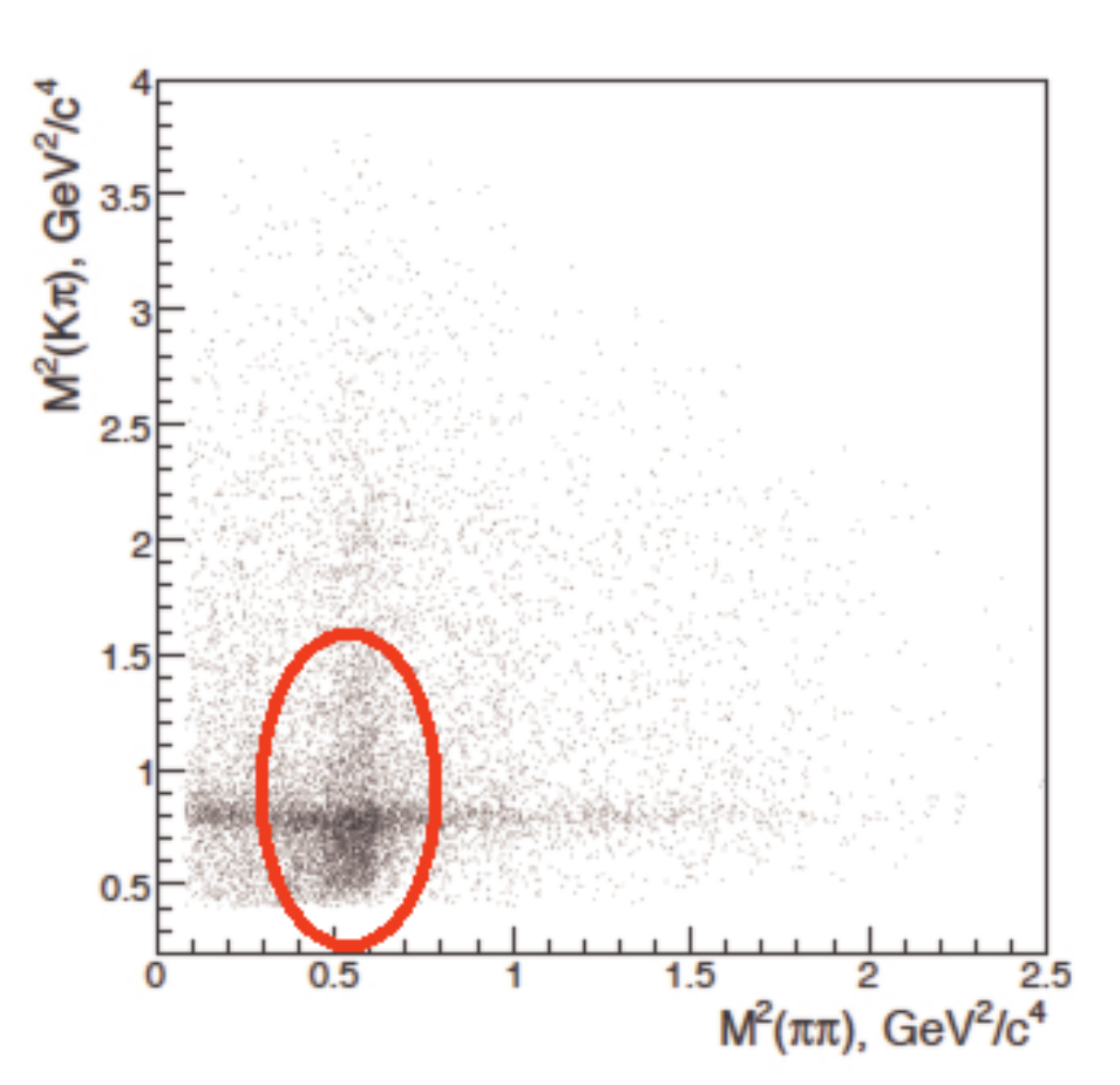}
		\includegraphics[width=0.3\textwidth]{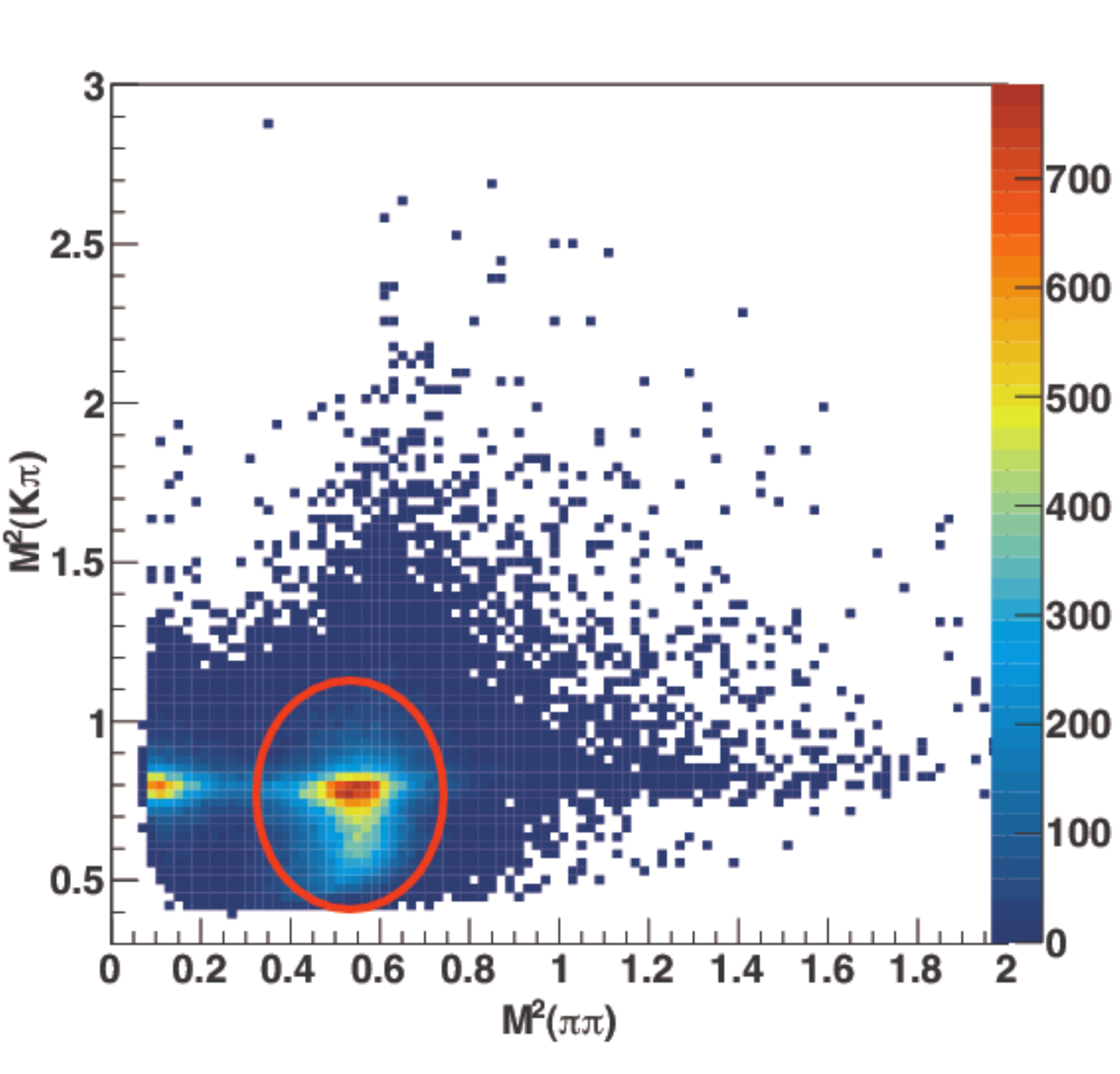}\hspace{3mm}
		\includegraphics[width=0.3\textwidth]{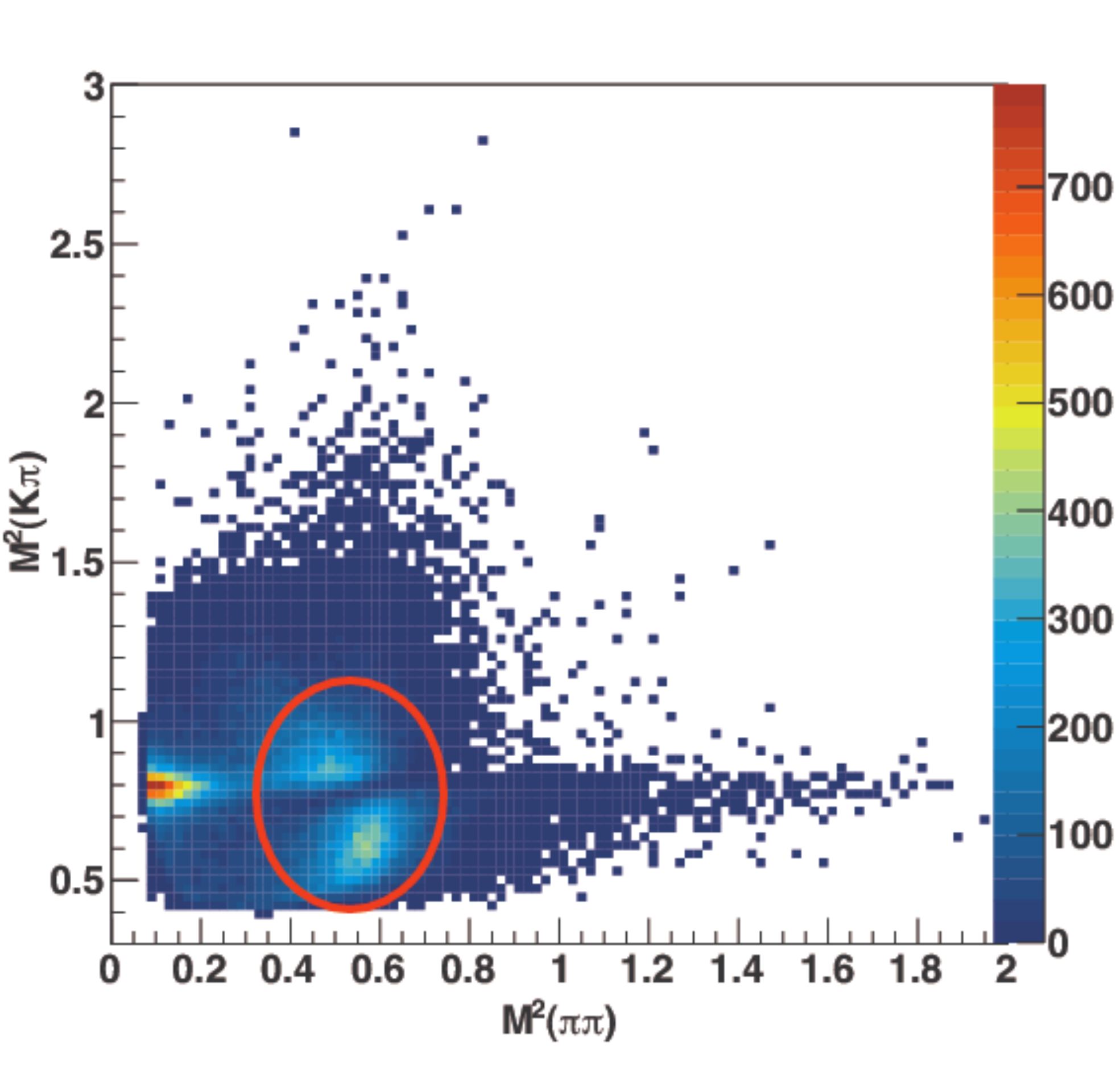}
	\caption{Dalitz plots of $B^+\to K_1^+(1270)\gamma\to K^+\pi^-\pi^+\gamma$, measured by the Belle collaboration~\cite{Belle:2010if} (left) and MC simulated for the ``offset'' phase equal to 0 (center) and $\pi$ (right) of the $K\rho$ channel relative to $(K^*\pi)_S$. The ``correct'' phase $\delta_\rho=0$ corresponds to our quark model prediction.}
	\label{fig:Dalitz_pipi-Kpi}
\end{figure}

In~\cite{Belle:2010if}, the Dalitz plots for $K_1(1270)\to K\pi\pi$ is shown in the three variable planes, $M^2(K^+\pi^+\pi^-)$, $M^2(K^+\pi^-)$ and $M^2(\pi^+\pi^-)$. On the Dalitz plot in the $M^2(K\pi)-M^2(\pi\pi)$ plane, a strong interference effect between $K_1\to K^*\pi$ and $K_1(1270)\to K\rho$ is observed (see Fig.~\ref{fig:Dalitz_pipi-Kpi}). In particular, it is pointed out that the weakening of the $K\rho$ in the region of $M(K\pi)>M_{K^*(892)}$ is originated from the interference of the $K\rho$ and $K^*\pi$ amplitudes. Here we will attempt to study the real phase (in another word, the relative sign) of the $K_1\to K^*\pi$ and $K_1(1270)\to K\rho$ amplitudes using this Dalitz plot, to check our theoretical prediction. 
Indeed, as we will see later-on, in a forthcoming paper, this information of the phase has an important consequence on our $\lambda_\gamma$ determination.   
%\begin{itemize}
%	\item According to Table~III only $S$ partial wave of the $K_1(1270)\to K^*\pi$ channel is taken into account.
%			\item A non negligible complex relative phase (apart from the complex phase of the Breit-Wigner factors) between $K\rho$ and $K^*\pi$ is measured: $(-44\pm4\pm7)$~deg
%	\end{itemize}

\subsubsection{Determining the relative sign of the $K\rho/K^*\pi$ amplitudes}

In this section, we demonstrate how the relative phase between the  $K\rho/K^*\pi$  amplitudes can be determined from the Dalitz plot. 

In~\cite{Belle:2010if}, the full amplitude of $K_1$ three-body decays is defined as

\begin{equation}
	|\mathcal{M}(s_{K_1},s_{K^*},s_\rho)|^2=\left|a_{K^*}A_{K^*}(s_{K_1},s_{K^*})+a_\rho A_\rho(s_{K_1},s_\rho) \right|^2
	\label{eq:amp2Belle}
\end{equation}
where the coefficients $a_{K^*,\,\rho}$ represent the strong decay of $K_1\to K\pi\pi$ through $K^*,\,\rho$ intermediate states. 
The amplitudes $A_{K^*,\,\rho}$ are defined as

\begin{equation}
	A_V(s_{K_1},s_V)=\frac{\sqrt{M_{K_1}\Gamma_{K_1}}}{M_{K_1}^2-s-iM_{K_1}\Gamma_{K_1}}\times\frac{\sqrt{M_V\Gamma_V}}{M_V^2-s_V-iM_V\Gamma_V}\times\sqrt{1+\frac{\vec{p}_i^2}{s_{K_1}}\cos^2\theta_{ik}}
	\label{eq:AJ1J2Belle}
\end{equation}
where $p_i$ is the breakup momentum of $P_i$ or $V$ in the $K_1$ reference frame and $\theta_{ik}$ is the angle between the momenta of $P_i$ and $P_k$ in the $V$ reference frame, which can be expressed in terms of $s_{K_1}$, $s_{ij}$, $s_{ik}$~\footnote{One has to notice that the $D$-wave amplitude is not taken into account in this parametrization and that the last factor in Eq.~\eqref{eq:AJ1J2Belle} corresponds to the $S$-wave.}.

Compared to the obtained Dalitz plot, we can determine the coefficients $a_{K^*,\,\rho}$ including the relative phase between them. The obtained result by the Belle collaboration yields~\cite{Belle:2010if}:

\begin{eqnarray}
	&|a_{K^*}|=0.962\pm0.058\pm0.176, \quad |a_\rho|=1.813\pm0.090\pm0.243& \nonumber \\
	&\delta_\rho\equiv\arg(a_\rho/a_{K^*})=-(43.8\pm4.0\pm7.3)^\circ& 
\label{eq:phase}
\end{eqnarray}
Formula~\eqref{eq:amp2Belle} can be written in the following general form factorizing out the phase:

\begin{equation}
	|\mathcal{M}(s_{K_1},s_{K^*},s_\rho)|^2=c_0(s_{K_1},s_{K^*},s_\rho)+c_1(s_{K_1},s_{K^*},s_\rho)\cos\delta_\rho+c_2(s_{K_1},s_{K^*},s_\rho)\sin\delta_\rho
	\label{}
\end{equation}
where $c_i(s,s_{K\pi},s_{\pi\pi})$ are the known functions, expressed in terms of various combinations of the real and imaginary parts of $|a_{K^*}|A_{K^*}(s_{K_1},s_{K^*})$ and $|a_\rho|A_\rho(s_{K_1},s_\rho)$. So, in order to establish the correspondence between our parametrization of $|\mathcal{M}|^2$ ($|\vec{\mathcal{J}}|^2$ in our case) one can compare the relative signs of the $\cos\delta_\rho$ and $\sin\delta_\rho$ coefficients, $c_{1,2}$, on the Dalitz plot. Direct numerical calculation shows that

\begin{equation}
	\text{sign}\left(c_1^\text{model}\right)=\text{sign}\left(c_1^\text{Belle}\right), ~ ~ ~ \text{sign}\left(c_2^\text{model}\right)=-\text{sign}\left(c_2^\text{Belle}\right)
	\label{eq:signKrhophase}
\end{equation}

\subsubsection{The issues of complex ``offset'' phases}

In principle, the QPCM predicts real $K_1\to VP$ amplitudes, without any complex phases. This should correspond to the $K$-matrix couplings. The complex rotation of the $K$-matrix states to the physical states should introduce complex phases but we found by explicit calculation that the imaginary part of the rotation angle is small:

\begin{equation}
	\varphi_{a^\prime\to a^{ph}}\simeq10^\circ
	\label{}
\end{equation}

However, the Belle collaboration measured a sizebly larger imaginary relative phase (i.e. Eq.~\eqref{eq:phase}) of $\delta_\rho\simeq-44^\circ$. We recall also that Daum {\it et al.} measured a non-zero phase of the order of 30$^\circ$. Similar value was found in the reanalysis of the ACCMOR data by the BABAR collaboration: $\delta_\rho=-31^\circ$ \cite{Babar:2009ii}.

There is no explanation of this complex phase in a definite theoretical model: neither in the $^3P_0$ quark model nor in the most general quasi-two-body $K$-matrix approach. Indeed, the ``offset'' phase which is introduced in the analysis by Daum {\it et al.} depends only on the decay channel and is the same for the lower and upper resonances. The general production amplitude for each channel in the reaction $K^-p\to(K^-\pi^+\pi^-)$ is written as \cite{Daum:1981hb,Babar:2009ii}

\begin{equation}
	F_i=e^{i\delta_i}\sum_j(1-iK\rho)_{ij}^{-1}P_j
	\label{eq:Amp_tot_Daum}
\end{equation}
where the factor $(1-iK)^{-1}$ represents the propagation and the decay of the $K_1$-resonance. The last factor $P_j$ describes the resonance production which can be in principle complex (indeed, one finds in \cite{Daum:1981hb} that there is a non-zero relative phase between the production couplings of two $K_1$-resonances). From Eq.~\eqref{eq:Amp_tot_Daum} it is obvious the ``offset'' phase $\delta_i$ can not be ascribed to either the resonance decay or production amplitude.

This puzzling situation must not be ignored and has to be studied more carefully. In the present, we use the model prediction for the $\mathcal{J}$-function as it is with pure real couplings. On the other hand, to adopt pragmatic attitude we explore the effect of the introducing this additional ``offset'' phase $\delta_\rho=-\delta_\rho^\text{Belle}$ in the calculation of the $\mathcal{J}$-function and the estimation of the theoretical uncertainty of $\lambda_\gamma$.

\subsection{The issue of the $\kappa\pi$ channel}

The PDG assigns a large branching ratio $\mathcal{B}(K_1(1270)\to K_0^*(1430)\pi)=(28\pm4)\%$. It is  extracted as all the branching ratios, from the ACCMOR data and analysis. However, this interpretation is dubious. The original ACCMOR measurement shows indeed a clear, strongly coupled peak in the ``scalar + $\pi$'' channel around the mass $M_{K\pi\pi}\sim1270$~MeV/$c^2$. However, it is not at all claimed that the scalar is $K_0^*(1430)$; it is treated as a lower and much broader scalar meson ($M\simeq1.25$~GeV/$c^2$, $\Gamma\simeq600$~MeV/$c^2$); or could be a continuum $(K\pi)_{S-\text{wave}}$ according to~\cite{Dunwoodie}.

\begin{figure}[h!]\centering
	\includegraphics[width=0.4\textwidth]{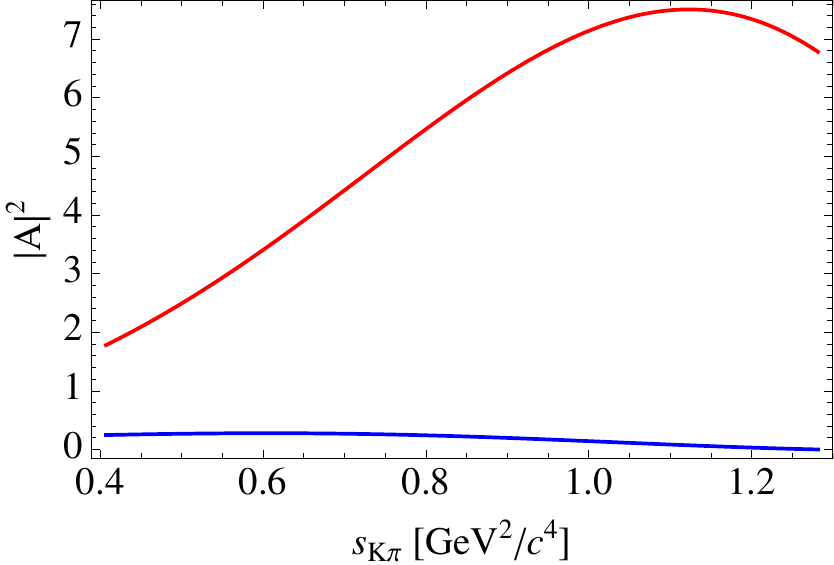}
	\caption{$|A_S(K_1(1270)\to K^*\pi)|^2$ (red) and $|A_P(K_1(1270)\to K_0^*(1430)\pi)|^2$ (blue) for $s_{K_1}=M_{K_1(1270)}^2$. The $K_1$ mixing angle $\theta_{K_1}$ is taken to be $60^\circ$.}
	\label{fig:K0st1430amplitude}
\end{figure}

The $K_0^*(1430)$-meson is the scalar orbitally excited state of kaon which has the mass $M_{K_0^*(1430)}=(1425\pm50)$~MeV/$c^2$ and width $\Gamma_{K_0^*(1430)}=(270\pm80)$~MeV/$c^2$ \cite{PDG}. According to quark models, the constituent quarks are in the $^3P_0$ state. In order to estimate the $K_0^*(1430)\pi$ contribution we use QPCM to calculate the $P$-wave amplitude for the decays $K_1(1270)\to K_0^*(1430)\pi$. One can see from Fig.~\ref{fig:K0st1430amplitude} that $A_P(K_1(1270)\to K_0^*(1430)\pi)$ is strongly suppressed compared to $A_S(K_1(1270)\to K^*\pi)$. Moreover, there is also a suppression due to the phase space. Finally, after the integration over the phase space for $\sqrt{s_{K\pi}}$ within the allowed physical range $[m_K+m_\pi;M_{K_1(1270)}-m_\pi]$, we predict that

\begin{equation}
	\frac{\mathcal{B}(K_1(1270)\to K_0^*(1430)\pi)}{\mathcal{B}(K_1(1270)\to K^*(892)\pi)}<0.01\%
	\label{}
\end{equation}
in blatant contradiction with the PDG entry.

What is most striking is that indeed, the Belle collaboration finds $\mathcal{B}(K_1(1270)\to K_0^*(1430)\pi)\simeq2\%$ (see Table~\ref{tab:BR_Belle}); it is very small as we predict. They did not find any other ``lower scalar+$\pi$'' component in the $K_1$-decay: the $\mathcal{B}$ missing with respect to ACCMOR seems to be filled by an enlargement of $K\rho$. Therefore, in our analysis, we do not include the  $K_1(1270)\to K_0^*(1430)\pi$ channel. Neither do we include any other possible scalar in the presented results. However, to take into account the contrary conclusions of ACCMOR, we keep in mind the possibility that there is some significant portion of the branching ratio carried by a very wide scalar meson, different from the $K_0^*(1430)$, such as the low lying state $K_0^*(800)$ (also called $\kappa$) \cite{DescotesGenon:2006uk}. Note that such state is most probably not a $q\overline{q}$ state and therefore the decay into $\kappa\pi$ can not be estimated within our theoretical model. Such contribution has not been tested explicitly in the analysis by the Belle collaboration.

Let us mention two other relevant facts: on the one hand the non-strange counter part of $\kappa(800)$, $\sigma$, is found with sizable branching ratio in the decay of $a_1(1260)$ in the $\sigma\pi$ state. On the other hand it is surprising, as noticed by Daum {\it et al.}, that there is no $\kappa\pi$ channel in the $K_1(1400)$-decay.

\section{Conclusions}\label{sec:conclusions}

Let us now summarise the main conclusions of the present work, and sketch some prospects for progress regarding theory as well as experience.

Not only the strong decay pattern of $K_1$-mesons is quite complex, but, not surprisingly, it is then difficult to analyze the whole system experimentally. In spite of many efforts, we have found that much  information is lacking, and that certain weaknesses may be suspected in various analyses. In lack of more fundamental treatments, we have recoursed to the quark model approach to explain and complement the experimental results. The quark model, although approximate, is the basis of our whole understanding of spectroscopy. The $^3P_0$ model for decays presents the advantage of handling in a simple way the whole set of $L=1$ decays~\footnote{In fact it is supported by a much larger set of experimental tests.}. On the other hand, experimental input is still required to fix necessary parameters, for instance the $K_1$ mixing angle. 

Our predictions for the specific strong decays under concern, i.e. $L=1$ states decaying to $VP$ states, can be evaluated by comparing to data where available. On the whole, our conclusion is very encouraging. In addition to the known fact  that a certain mixing is able to explain the pattern of $VP$ decays, the model explains detailed features which are quite outside an $SU(3)$ symmetry approach, and require a \underline{\it dynamical} approach. This is the case, for instance, the $D/F$ ratio of octet couplings and the $D/S$ ratios in magnitude and \underline{\it phase}; it is an achievement of the model that all the observables phases are correctly predicted. Another typically dynamical prediction is that the decays to $q\bar{q}$ scalar+pseudoscalar should be very small. It is in agreement with the recent observations by Belle~\cite{Belle:2010if}, but it does not exclude a large contribution of non $q\bar{q}$ scalars, which could then explain the observations of ACCMOR. In any case, it strengthens the conclusion that the $\kappa$ channel observed by ACCMOR is not the $K_0^*(1430)$ (presumed $q\bar{q}$) as tabulated in the PDG tables.

We believe that the $K_1$ system deserves further investigation because it has revealed interesting in various aspects. Indeed, it also presents unexplained features in the standard domain of spectroscopy, i.e. the mixing angle and the mass splitting. The mixing of the two states offers the possibility to explain the remarkable pattern of $K^*\pi/K\rho$ decays, but the angle is not a theoretical prediction. In fact, in a potential model, spin-orbit forces generate a mixing, but it is not the one which is observed. As explained in the text, loop effects would also generate a mixing effect, but it cannot be calculated. It must also be noted that the mixing does not explain why the $\kappa\pi$ channel if present in $K_1(1270)$ is absent in $K_1(1400)$: quite on the contrary, as explained in the paper of Daum {\it et al.}, one would expect the mixing to generate a coupling from the $K_A$ component. Apart from mixing, the predictions of potential  model also fail to explain the splitting of the two states - it is predicted much too small by the model of  Godfrey and Isgur, which can be estimated to be the most trustable. These facts show that our knowledge of spectroscopy is not yet satisfactory even for apparently well identified, low-lying $q\bar{q}$ states. Understanding these facts then justify further studies. 
\vspace*{0.4cm}

\noindent
We try to say something about possible improvements:

\vspace*{0.2cm}
\noindent
1 \underline{\it Improvement of the theoretical treatment} \\
It is important to recall that there is no fundamental theoretical treatment of such problems and that quark models, on which our theoretical model is based, although much valuable, contain essential approximations, i.e. ones  that cannot be improved systematically. This holds in two respects: potential models are of course essentially approximate, even with relativistic improvements as included in the model of  Godfrey and Isgur, but this is also true of the quark-pair-creation decay model itself, which, presently, is essentially non-relativistic. The center of mass motion of the hadrons is not treated relativistically. Progress is desirable in this direction.

\vspace*{0.4cm}

\noindent
2 \underline{\it Prospects of improvement of experimental knowledge} \\
At present, further progress could come mainly from a better and more complete determination of the magnitudes and the phases of the various couplings by experiments. Certainly, the old experiments with production of $K_1$ by strong interaction scattering, as the ones of SLAC and ACCMOR, have much larger statistics for decays involving $K_1$ than present B factories . Yet there is little prospect of them being redone, and they have also their own weakness in the fact that the production process is complex. On the other hand, there is the hope that new detailed studies could be made in $B$ and $\tau$-decays. Encouraging examples have been coming from both BABAR and Belle, such as $\tau\to K_1\nu_\tau$~\cite{Belle:2010tc} and $B\to K_1\psi$~\cite{Belle:2010if}. In fact, a distribution in an additional angle may also help to improve the analysis. For example, a new study of $B\to K_1\psi$ with angular analysis could yield directly the crucial quantity $Im(\vec{n}\cdot(\vec{\mathcal{J}}\times\vec{\mathcal{J}^*}))$ up to a multiplicative constant~\cite{Kou:2010kn}. The analyses could be guided by our semi-theoretical and approximate investigation, which, for instance, emphasizes the need to take into account $D$ waves, not included in the present Belle analysis of $B\to K_1\psi$.

\vskip15mm

\section*{Acknowledgments}
We thank very much Damir Becirevic for his critical discussions and comments, and for constant help. Alain Le Yaouanc acknowledges constant discussions with his coworkers Luis Oliver and Jean-Claude Raynal. A.T. thanks Olivier P\`ene for his precious help. We would like to thank W.~Dunwoodie, S.~\'Emery, Y.~Sakai, K.~Trabelsi, M.~Nakao and S.~Hashimoto for very useful discussions and provided information. This work was supported in part by the ANR contract ``LFV-CPV-LHC'' ANR-NT09-508531 and France-Japan corporation of IN2P3/CNRS TYL-LIA.

\newpage

\appendix

\section{Re-interpreting the ACCMOR result in terms of ``physical states``}\label{app:ACCMOR}

In order to determine our model parameters and the $K_1$ mixing angle from comparison of the predicted partial decay widths of the $K_1$-meson decays into the dominant $K^*\pi$ and $K\rho$ channels with the measured experimental values, we use the fitted $K$-matrix parameters extracted by Daum {\it et al.} from Ref.~\cite{Daum:1981hb} (see Table~\ref{tab:Daum2}).

Using the definition of the $K_1$ mixing by Daum {\it et al.}
(which is different from \eqref{eq:K1_mixing} that we use by signs):

\begin{equation}
	\begin{split}
		|K_1(1400)\rangle &= |K_{1A}\rangle\cos\theta_{K_1}+|K_{1B}\rangle\sin\theta_{K_1} \\
		|K_1(1270)\rangle &= -|K_{1A}\rangle\sin\theta_{K_1}+|K_{1B}\rangle\cos\theta_{K_1}
	\end{split}
	\label{eq:K1_mixing_Daum}
\end{equation}
the dominant $S$-wave $K$-matrix couplings of the $K_1$'s to the states $K^*\pi$ (channel 1) and $K\rho$ (channel 2) are given as \cite{Daum:1981hb}

\begin{equation}
	\begin{split}
		f_{a^\prime1} &= \frac{1}{2}\gamma_+\cos\theta_{K_1}+\sqrt{\frac{9}{20}}\gamma_-\sin\theta_{K_1} \\
		f_{b^\prime1} &=-\frac{1}{2}\gamma_+\sin\theta_{K_1}+\sqrt{\frac{9}{20}}\gamma_-\cos\theta_{K_1} \\
		f_{a^\prime2} &= \frac{1}{2}\gamma_+\cos\theta_{K_1}-\sqrt{\frac{9}{20}}\gamma_-\sin\theta_{K_1} \\
		f_{b^\prime2} &=-\frac{1}{2}\gamma_+\sin\theta_{K_1}-\sqrt{\frac{9}{20}}\gamma_-\cos\theta_{K_1}
	\end{split}
	\label{eq:K-matrix_couplings_Daum}
\end{equation}
where $\gamma_+$ and $\gamma_-$ are the reduced $SU(3)$ couplings for $K_{1A}$ ($F$-type) and $K_{1B}$ ($D$-type) respectively. Their fitted experimental values are given in Table~\ref{tab:Daum1}. The indices $a^\prime$ and $b^\prime$ denote the upper and lower $K_1$ resonances.

\begin{table}[h]
	\centering
	\begin{tabular}{|c|c|c|c|c|c|c|c|}
		\hline
		$m_{a^\prime}$ , GeV/$c^2$ & $m_{b^\prime}$ , GeV/$c^2$ & $\gamma_+$ & $\gamma_-$ & $\tilde\theta_{K_1}$ \\
		\hline
		1.4$\pm$0.02 & 1.17$\pm$0.02 & 0.78$\pm$0.1 & 0.54$\pm$0.1 & 64$^\circ\pm$8$^\circ$\\
		\hline
	\end{tabular}
	\caption{Fitted $K$-matrix pole masses, $S$-wave reduced $SU(3)$ couplings and mixing angle for $K_{1A}$ ($F$-type) and $K_{1B}$ ($D$-type), taken from Ref.~\cite{Daum:1981hb} (low $t$ data). The indices $a^\prime$ and $b^\prime$ denote the upper and lower $K_1$ resonances.}
	\label{tab:Daum1}
\end{table}
\begin{table}[h]
	\centering
	\begin{tabular}{|c|c|c|c|c|c|}
		\hline
		$f_{a^\prime1}$ & $f_{b^\prime1}$ & $f_{a^\prime2}$ & $f_{b^\prime2}$ & $f_{a^\prime3}$ & $f_{b^\prime3}$ \\
		\hline
		0.50$\pm$0.07 & -0.19$\pm$0.09 & -0.15$\pm$0.10 & -0.51$\pm$0.06 & 0 & 0.32 \\
		\hline
	\end{tabular}
	\caption{$K$-matrix couplings, calculated from Eq.~\eqref{eq:K-matrix_couplings_Daum} using the fitted parameters from Table~\ref{tab:Daum1}. The indices $a^\prime$ and $b^\prime$ denote the upper and lower $K_1$ resonances decaying into $K^*\pi$ (channel 1) and $K\rho$ (channel 2) hadronic states respectively. The coupling to the $K_0^*(1430)\pi$ channel, where $K_0^*(1430)$ resonance is supposed to have the mass 1.25~GeV/$c^2$ and width 600~MeV/$c^2$, $f_{b3}$ is taken from Ref.~\cite{Babar:2009ii}.}
	\label{tab:Daum2}
\end{table}

Using the experimental values of the $K$-matrix couplings from Table~\ref{tab:Daum2} and performing the diagonalization of the complex mass matrix \eqref{eq:K-matrix_mass_matrix}, we observed that 

\begin{figure}[hp!]
	\begin{center}
		\includegraphics[width=0.45\textwidth]{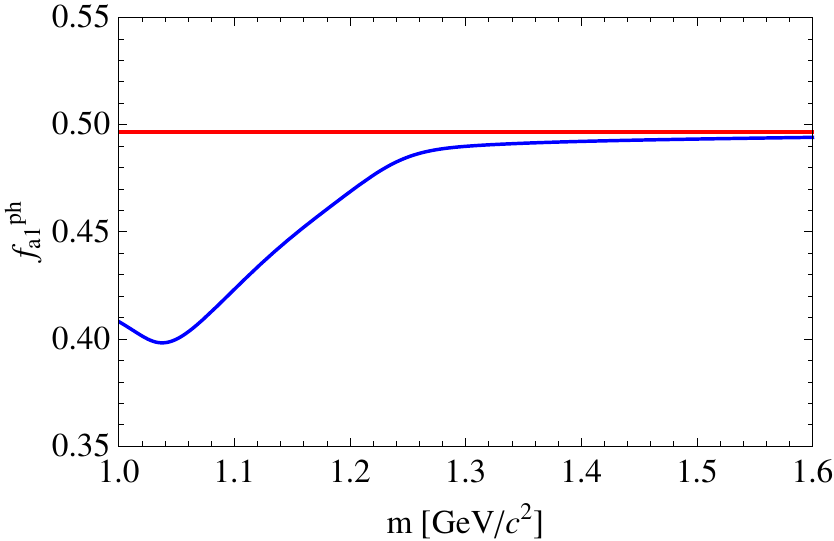}\hspace{5mm}
		\includegraphics[width=0.45\textwidth]{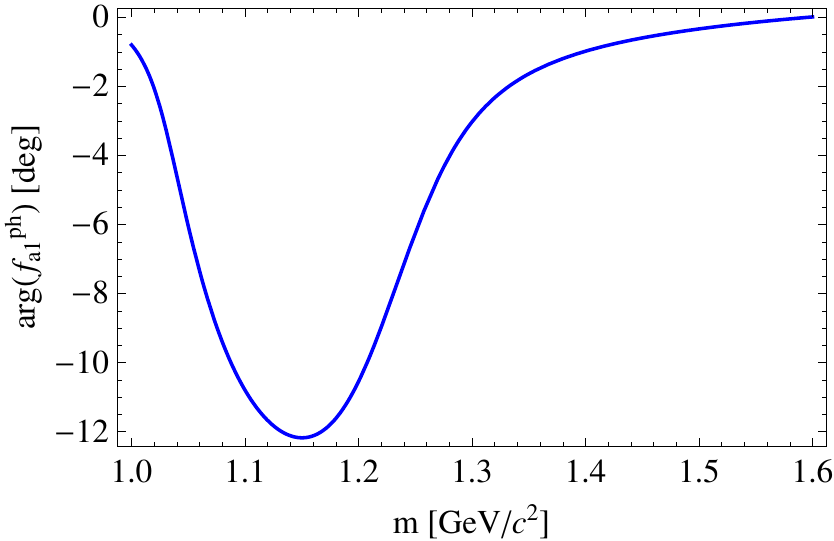}\hspace{5mm}
		\includegraphics[width=0.45\textwidth]{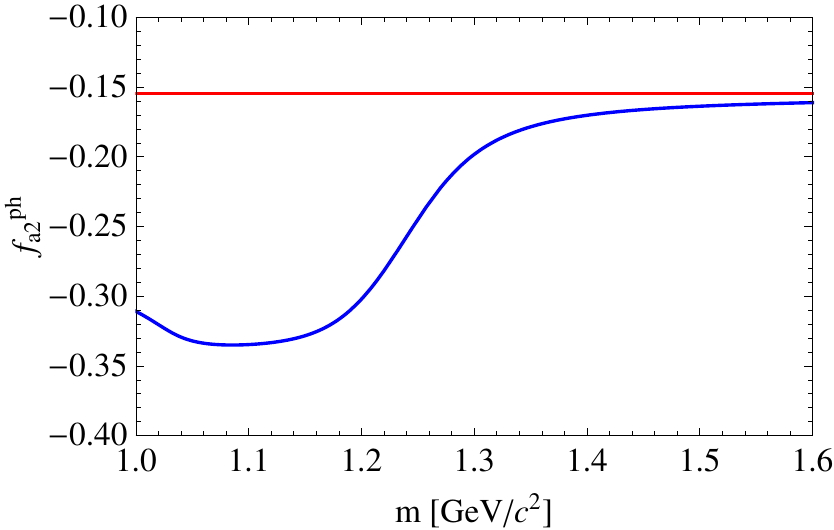}\hspace{5mm}
		\includegraphics[width=0.45\textwidth]{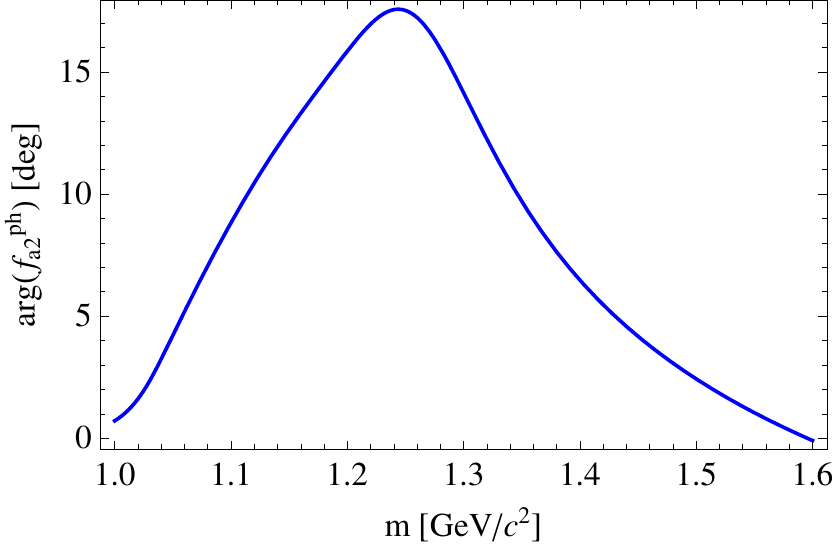}\hspace{5mm}
		\includegraphics[width=0.45\textwidth]{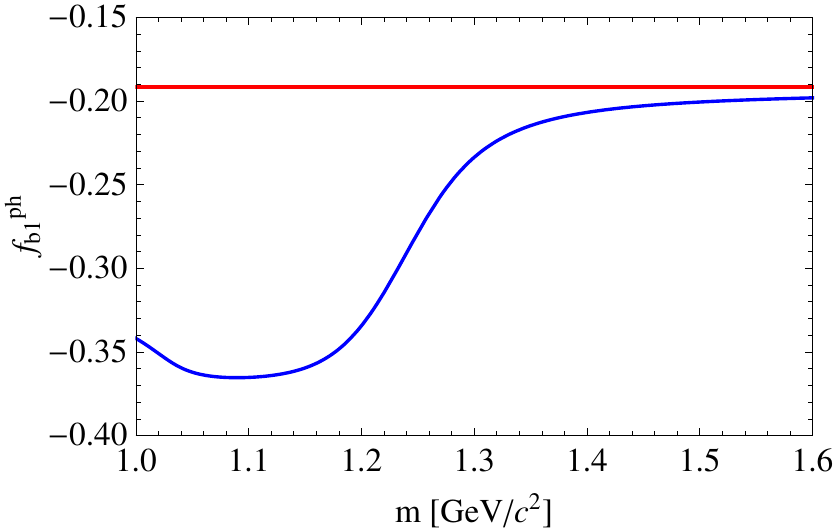}\hspace{5mm}
		\includegraphics[width=0.45\textwidth]{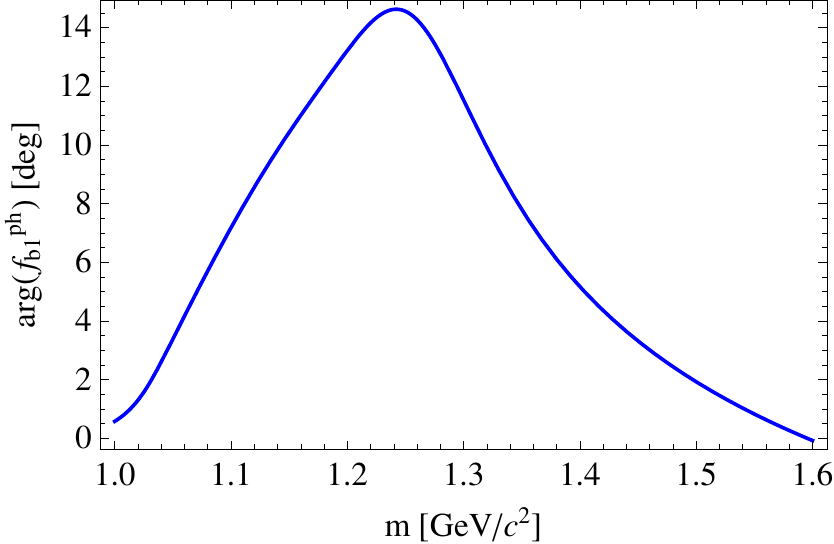}\hspace{5mm}
		\includegraphics[width=0.45\textwidth]{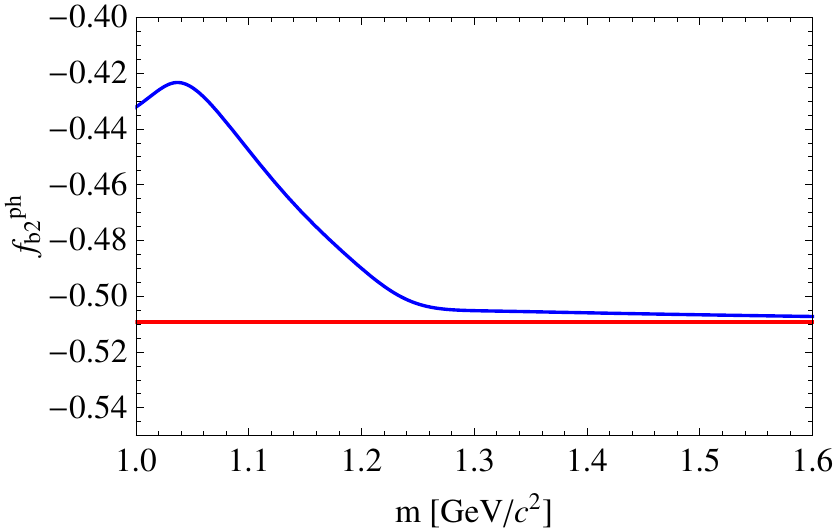}\hspace{5mm}
		\includegraphics[width=0.45\textwidth]{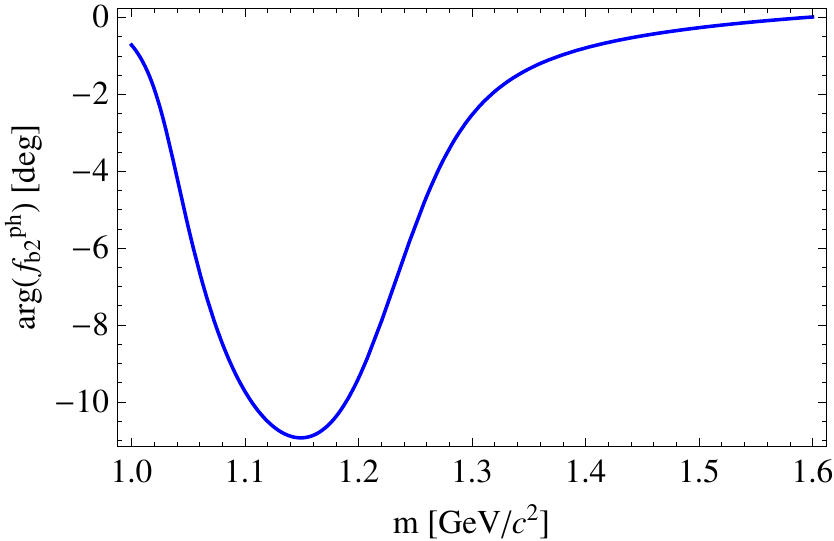}
	\end{center}
	\caption{Energy dependence of the physical couplings (blue). The red lines represent the values of the real couplings for the $K$-matrix states, fitted by ACCMOR collaboration \cite{Daum:1981hb}.}
	\label{fig:rotated_couplings}
\end{figure}

\begin{figure}[h!]
	\begin{center}
		\includegraphics[width=0.45\textwidth]{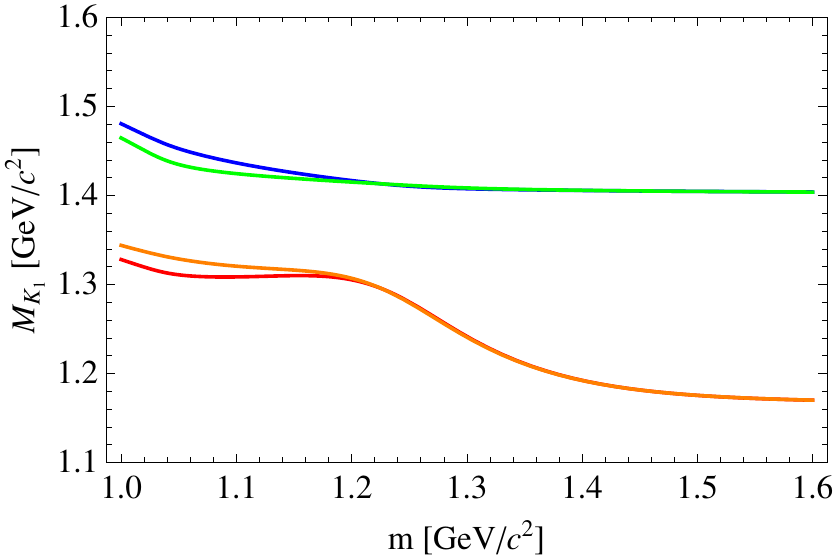}\hspace{5mm}
		\includegraphics[width=0.45\textwidth]{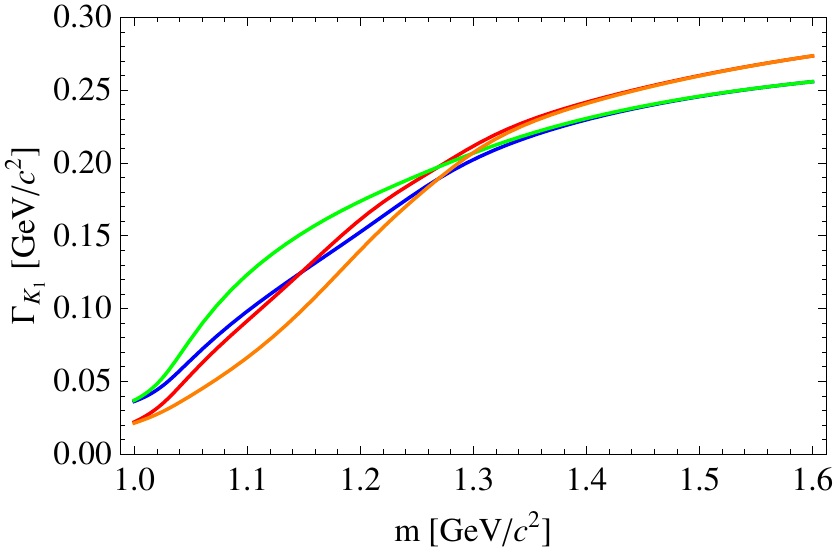}
	\end{center}
	\caption{Energy dependence of the mass (left) and width (right) of $K_1(1270)$ (red, orange) and $K_1(1400)$ (blue, green). Red and blue curves correspond to the masses and total widths of the physical eigenstates, i.e. diagonal mass matrix elements \eqref{eq:physical_pole_masses} which are calculated in terms of the rotated physical couplings. Orange and green curves represent the leading diagonal elements of the complex mass matrix $\left(M^\prime -i\Gamma^\prime/2\right)_{a^\prime b^\prime}$ \eqref{eq:Sprime2} in the $K$-matrix eigenstate basis. The $D$-wave contribution is not taken into account due to the absence of knowledge of the corresponding couplings.}
	\label{fig:Mass_Width_diag}
\end{figure}

\begin{itemize}
	\item The variation of the absolute values and phases of the new rotated physical couplings $\{f_{a^{ph}i},f_{b^{ph}i}\}$ around the masses at the peak of Breit-Wigner (i.e. $m\sim$1.27~GeV/$c^2$ and 1.4~GeV/$c^2$) turn out to be small (see Fig.~\ref{fig:rotated_couplings}).
	\item Contribution of the complex phase space for energy below the decay threshold (which implies $\rho_{ij}(m)\to i|\rho_{ij}(m)|$) is very small for diagonalized physical mass of $K_1(1400)$ (see Fig.~\ref{fig:Mass_Width_diag}). But one observes a threshold effect for $K_1(1270)$ near $m\sim1.2$~GeV/$c^2$. However, the mass variation of $M_{K_1}(m)$ around the peak of Breit-Wigner can be considered not so significant.
	\item One can see from Fig.~\ref{fig:Mass_Width_diag} that, contrary to $M_{K_1}(m)$ dependence, the width $\Gamma_{K_1}(m)$ is a rapidly varying function of the energy $m$.
	\item Non-diagonal elements of the mass matrix \eqref{eq:K-matrix_mass_matrix} are sufficiently small compared to the diagonal ones. One can see from Fig.~\ref{fig:Mass_Width_diag} that the difference between the properly diagonalized masses and widths (blue/red curves), which are calculated in terms of the rotated physical couplings, and the real and imaginary parts of the diagonal elements of \eqref{eq:K-matrix_mass_matrix} (green/orange curves) is insignificant. As a consequence,  our assumption for the partial widths

		\begin{equation}
			\Gamma_{a^{ph}i}(M_{peak})\simeq\Gamma_{a^\prime i}(M_{peak})=\Gamma_{a^\prime i}^{QPCM}(M_{peak})
			\label{}
		\end{equation}
		seems to be reasonable. This means that we can use the experimental measured $K$-matrix couplings in order to calculate the partial decay widths and fit our moedel parameters, namely quark-pair-creation constant $\gamma$ and the mixing angle $\theta_{K_1}$, which can further be used for the $\mathcal{J}$ function computation.
\end{itemize}

\section{QPCM}\label{app:QPCM}

\subsection{Spacial integrals in QPCM}

For the decay $A\to B+C$ (see Fig.~\ref{fig:QPCpic}) the spacial integrals are given by

\begin{equation}
	\begin{split}
		I_m^{(ABC)} =& \int d^3\vec{k}_1d^3\vec{k}_2d^3\vec{k}_3d^3\vec{k}_4\delta(\vec{k}_1+\vec{k}_2-\vec{k}_A)\delta(\vec{k}_2+\vec{k}_3-\vec{k}_B)\delta(\vec{k}_4+\vec{k}_1-\vec{k}_C)\delta(\vec{k}_3+\vec{k}_4) \\
	& \times\mathcal{Y}_1^m(\vec{k}_3-\vec{k}_4)\psi^{(A)}(\vec{k}_1-\vec{k}_2)\psi^{(B)}(\vec{k}_2-\vec{k}_3)\psi^{(C)}(\vec{k}_4-\vec{k}_1) \\
	=& \frac{1}{8}\int d^3\vec{k}\mathcal{Y}_1^m(\vec{k}_B-\vec{k})\psi^{(A)}(\vec{k}_B+\vec{k})\psi^{(B)}(-\vec{k})\psi^{(C)}(\vec{k})
	\label{}
	\end{split}
\end{equation}
where $\psi$'s are the normalized Fourier transforms of harmonic oscillator meson wave functions. The wave functions for the ground ($L=0$) and orbitally excited ($L=1$) meson states are defined as
\begin{equation}
	\begin{split}
	\psi_0^{(i)}(\vec{k})   &= \frac{R_i^{3/2}}{\pi^{3/4}}\exp\left(-\frac{\vec{k}^2R_i^2}{8}\right)~ ~ ~ ~ ~ ~ ~ ~ ~ ~ ~ ~ ~ ~ ~ ~ ~ ~ ~ ~ ~ ~ (L=0) \\
	\psi_1^{m(i)}(\vec{k}) &= \sqrt{\frac{2}{3}}\frac{R_i^{5/2}}{\pi^{1/4}}\mathcal{Y}_1^m(\vec{k})\exp\left(-\frac{\vec{k}^2R_i^2}{8}\right) ~ ~ ~ (L=1,L_z=m) \\
	\mathcal{Y}_1^m(\vec{k}) &= |\vec{k}|Y_1^m(\hat{\vec{k}})=(\vec\varepsilon_m\vec{k})\sqrt{\frac{3}{4\pi}}
	\label{}
\end{split}
\end{equation}
Here $R_i$ is the meson wave function radius and $\vec{\varepsilon}_m$ are the $A$-polarization vectors, defined as
\begin{equation}
	\vec{\varepsilon}_0=\left(\begin{array}{ccc}0\\0\\1\end{array}\right),~ ~ ~ ~ \vec{\varepsilon}_{\pm1}=\mp\frac{1}{\sqrt2}\left(\begin{array}{ccc}1\\\mp i\\0\end{array}\right)
	\label{}
\end{equation}

Performing the integration over $\vec{k}$ one obtains for the orbitally excited axial-vector meson decay into pseudoscalar and vector mesons in the $A$-meson reference frame:
\begin{equation}
	\begin{split}
		I_m^{(ABC)} = -\frac{4\sqrt3}{\pi^{5/4}}\frac{R_A^{5/2}(R_B R_C)^{3/2}}{(R_A^2+R_B^2+R_C^2)^{5/2}}\left((\vec{\varepsilon}_m\cdot\vec{\varepsilon}_{-m})-(\vec{\varepsilon}_m\cdot\vec{k}_B)(\vec{\varepsilon}_{-m}\cdot\vec{k}_B) \right. \\
		 \left.\times\frac{(2R_A^2+R_B^2+R_C^2)(R_B^2+R_C^2)}{4(R_A^2+R_B^2+R_C^2)}\right)\exp\left[-\vec{k}_B^2\frac{R_A^2(R_B^2+R_C^2)}{8(R_A^2+R_B^2+R_C^2)}\right]
	\label{}
	\end{split}
\end{equation}
Setting $\vec{k}_B$ along $z$-axis, the integrals become
\begin{equation}
	\begin{split}
	I_0^{(ABC)} =& -\frac{4\sqrt3}{\pi^{5/4}}\frac{R_A^{5/2}(R_B R_C)^{3/2}}{(R_A^2+R_B^2+R_C^2)^{5/2}}\left(1-\vec{k}_B^2\frac{(2R_A^2+R_B^2+R_C^2)(R_B^2+R_C^2)}{4(R_A^2+R_B^2+R_C^2)}\right) \\
	& \times \exp\left[-\vec{k}_B^2\frac{R_A^2(R_B^2+R_C^2)}{8(R_A^2+R_B^2+R_C^2)}\right] \\
	I_1^{(ABC)} =& \frac{4\sqrt3}{\pi^{5/4}}\frac{R_A^{5/2}(R_B R_C)^{3/2}}{(R_A^2+R_B^2+R_C^2)^{5/2}}\exp\left[-\vec{k}_B^2\frac{R_A^2(R_B^2+R_C^2)}{8(R_A^2+R_B^2+R_C^2)}\right]
	\label{eq:ImAVP}
	\end{split}
\end{equation}

For the vector meson ground state decay into two pseudoscalar mesons the spacial integral is
\begin{equation}
	\begin{split}
		I_m^{(ABC)} = \frac{\sqrt6}{\pi^{5/4}}(\vec{\varepsilon}_m\cdot\vec{k}_C)\frac{(R_A R_B R_C)^{3/2}(2R_A^2+R_B^2+R_C^2)}{(R_A^2+R_B^2+R_C^2)^{5/2}} \\
		\times\exp\left[-\vec{k}_C^2\frac{R_A^2(R_B^2+R_C^2)}{8(R_A^2+R_B^2+R_C^2)}\right]
	\label{eq:ImVPP}
	\end{split}
\end{equation}

\subsection{Fixing the relative signs for three-body decay} \label{signs}

\subsubsection{Clebsch-Gordan coefficients}

%We decided to use the convention ``down$\rightarrow$up'' (see Fig.~\ref{fig:K1toKpipi}). 
As it was emphasized in the text the relative sign of several amplitudes involving various intermediate states plays very important role. Therefore the convention of the particle order in the Clebsch-Gordan coefficients is very important. For instance, for the case of $K_1^+\to K^+\pi^-\pi^+$ decay that implies that we take the Clebsch-Gordan coefficients defined in the following way:

\begin{equation}
	\begin{split}
	(K^{*0}\pi^+|K_1^+) &= (1/2,-1/2;1,1|1/2,1/2)=-\sqrt{\frac{2}{3}} \\
	(K^+\pi^-|K^{*0}) &= (1/2,1/2;-1,1|-1/2,1/2)=+\sqrt{\frac{2}{3}} \\
	(K^+\rho^0|K_1^+) &= (1/2,1/2;1,0|1/2,1/2)=+\frac{1}{\sqrt{3}} \\
	(\pi^-\pi^+|\rho^0) &= (1,-1;1,1|1,0)=-\frac{1}{\sqrt{2}}
	\end{split}
	\label{eq:ClebschGordan}
\end{equation}
This gives the signs of the amplitudes listed in Table~\ref{tab:K1ABamplitudes}.% and the coefficients $-2/3$ for $K^*\pi$ and $-1/\sqrt{6}$ for $\rho K$ contributions in Eq.~24 from Ref.~\cite{Gronau:2002rz}.

\subsection{Determination of the relative sign of $g_{K^*K\pi}$ and $g_{\rho\pi\pi}$}

Following the definition in the work of Gronau {\it et al.}, the total amplitude of the two possible channels is written as

\begin{equation}
\begin{split}
	\mathcal{M}^{(a)} & = \varepsilon_\mu^{(K_1)}T_{K^*\pi}^{\mu\nu}\varepsilon_\nu^{(K^*)*}g_{K^*K\pi}\varepsilon_\sigma^{(K^*)}(p_{\pi^-}-p_{K^+})^\sigma \\
	\mathcal{M}^{(b)} & = \varepsilon_\mu^{(K_1)}T_{K\rho}^{\mu\nu}\varepsilon_\nu^{(\rho)*}g_{\rho\pi\pi}\varepsilon_\sigma^{(\rho)}(p_{\pi^+}-p_{\pi^-})^\sigma
\end{split}
	\label{}
\end{equation}
where $T_{VP}^{\mu\nu}$ is the hadronic tensor, parametrized in terms of the form factors $f_V$, $h_V$ (or equivalently the $S$ and $D$ partial wave amplitudes)~\footnote{For a more detailed definition of the hadronic tensor $T^{\mu\nu}$ and its parametrization in terms of two form factors, $f_V$ and $h_V$, see Ref.~\cite{Kou:2010kn}.}.

Now, using the same Clebsch-Gordan coefficients, defined above in Eq.~\eqref{eq:ClebschGordan}, one can write the amplitude of the $V\to PP$ decay, calculated the general tensor Lorenz-invariant form in the vector meson reference frame:

\begin{equation}
	\begin{split}
		\mathcal{M}(K^{*0}\to K^+\pi^-) & = -\sqrt{\frac{2}{3}}g_{K^*K\pi}(\vec{\varepsilon}_{K^*}\cdot(\vec{p}_{\pi^-}-\vec{p}_{K^+})) = \sqrt{\frac{8}{3}}g_{K^*K\pi}(\vec{\varepsilon}_{K^*}\cdot\vec{p}_{K^+}) \\
		\mathcal{M}(\rho^0\to \pi^-\pi^+) & = \sqrt{\frac{1}{2}}g_{\rho\pi\pi}(\vec{\varepsilon}_{\rho}\cdot(\vec{p}_{\pi^+}-\vec{p}_{\pi^-})) = -\sqrt2g_{\rho\pi\pi}(\vec{\varepsilon}_{\rho}\cdot\vec{p}_{\pi^-})
	\end{split}
	\label{}
\end{equation}

Taking into account all the spin and isospin couplings, the QPCM prediction is

\begin{equation}
	\begin{split}
	\mathcal{M}_m^{\text{QPCM}}(K^{*0}\to K^+\pi^-) & = -\frac{1}{6}\gamma I_m^{(K^*K\pi)}=-\frac{1}{6}\gamma \tilde{I}^{(K^*K\pi)}(\vec{\varepsilon}_m\cdot\vec{p}_{K^+}) \\
		\mathcal{M}_m^{\text{QPCM}}(\rho^0\to\pi^-\pi^+) & = -\frac{1}{3\sqrt2}\gamma I_m^{(\rho\pi\pi)}=-\frac{1}{3\sqrt2}\gamma \tilde{I}^{(\rho\pi\pi)}(\vec{\varepsilon}_m\cdot\vec{p}_{\pi^-})
	\end{split}
	\label{}
\end{equation}
where $\tilde{I}^{(VPP)}$ can be defined from Eq.~\eqref{eq:ImVPP}~\footnote{One has to be careful with the choice of the momentum, i.e. $\vec{p}_C$ or $\vec{p}_B=-\vec{p}_C$, since it changes the sign of the $P$-wave amplitude.}.

Now, doing a matching between two approaches and factorizing out the common factor $\vec{\varepsilon}\cdot\vec{p}_i$, we can write the following equations:

\begin{equation}
\begin{split}
	\sqrt{\frac{8}{3}}g_{K^*K\pi} & = -\frac{1}{6}\gamma \tilde{I}^{(K^*K\pi)} \\
	-\sqrt2 g_{\rho\pi\pi} & = -\frac{1}{3\sqrt2}\gamma \tilde{I}^{(\rho\pi\pi)}
\end{split}
	\label{}
\end{equation}
Since $\tilde{I}^{(VPP)}$ is a positive function, one can see that 
\begin{equation}
	\text{sign}(g_{K^*K\pi})=-\text{sign}(g_{\rho\pi\pi}) \nonumber
\end{equation}
and in the $SU(3)$ limit $\frac{g_{\rho\pi\pi}}{g_{K^*K\pi}}=-\sqrt{\frac{8}{3}}$.

One can notice that the choice of the order in the isospin factors of the vector meson decay into two pseudoscalars in Eq.~\eqref{eq:ClebschGordan} well fixes the relative sign of the $g_{VPP}$ couplings. Moreover, this method makes the calculation of the quasi-two-body decay amplitude independent on the intermediate vector meson state ($K^*$, $\rho$) wave function sign (which, in principle, can be arbitrary in the quasi-two-body calculation since the final state is not the same)!

\section{Partial Wave Amplitudes}

With the quark models one can directly calculate the amplitudes with definite spin or helicity states. An experiment can measure the partial wave amplitudes of particular quantum numbers of the final state. Since both canonical (orbital) and helicity approaches give complete description of the process, one can find the relation between two representations for the decay of the initial at-rest state $|J,M\rangle$ with spin $J$ and spin projection $M$ on to the $z$-axis into two particles with spins $s_{1,2}$, helicities $\lambda_{1,2}$, total spin $S$ and relative orbital momentum $L$ ~\cite{Chung:1971ri}:

\begin{equation}
	\mathcal{M}_{\lambda_1\lambda_2}^{JM}(\Omega_1) = N_Jf_{\lambda_1\lambda_2}^J D_{M,\lambda_1-\lambda_2}^{J*}(\Omega_1)
	\label{}
\end{equation}
with the normalization factor $N_J=\sqrt{\frac{2J+1}{4\pi}}$.

The observed number of events is given by

\begin{equation}
	\sum_{M,\lambda_i,\lambda_i^{'}}\int \mathcal{M}_{\lambda_1\lambda_2}^{JM}(\Omega_1) \mathcal{M}_{\lambda_1^{'}\lambda_2^{'}}^{JM*}(\Omega_1)d\Omega_1=4\pi\sum_{\substack{\lambda_i,\lambda_i^{'}\\\lambda_1-\lambda_2=\lambda_1^{'}-\lambda_2^{'}}} N_J^2f_{\lambda_1\lambda_2}^J f_{\lambda_1^{'}\lambda_2^{'}}^{J*}
	\label{}
\end{equation}

The recoupling from the canonical to the helicity representation is
\begin{equation}
	N_Jf_{\lambda_1\lambda_2}^J = \sum_{L,S}\sqrt{2L+1}(L,0;S,\lambda_1-\lambda_2|J,\lambda_1-\lambda_2)(s_1,\lambda_1;s_2,-\lambda_2|S,\lambda_1-\lambda_2)A_L
	\label{}
\end{equation}

The two-body decay of the axial-vector meson into vector and pseudoscalar mesons can proceed in $S$ and $D$-waves. Using $J=1$, $\lambda_1=\lambda_V$, $\lambda_2=0$, the helicity amplitudes in the $A$ reference frame can be written in terms of partial wave amplitudes:

\begin{equation}
	N_1f_{\lambda_V0}^1 = \sum_{L=0,2}\sqrt{2L+1}(L,0;1,\lambda_V|1,\lambda_V)A_L
	\label{}
\end{equation}

 Setting $\vec{k}_V$ along $z$-direction (i.e. $\theta_V=0$), the helicity amplitudes are

\begin{equation}
	\begin{split}
		\mathcal{M}_{00}^{10}=N_1f_{00}^{1} &= A_S-\sqrt{2}A_D \\
	\mathcal{M}_{\pm1,0}^{1,\pm1}=N_1f_{\pm1,0}^{1} &= A_S+\frac{1}{\sqrt{2}}A_D
	\end{split}
	\label{}
\end{equation}

By-turn, the partial wave amplitudes are related to the helicity amplitudes as following:

\begin{equation}
	\begin{split}
	A_S &= \frac{1}{3}(2\mathcal{M}_{10}^{11}+\mathcal{M}_{00}^{10}) \\
	A_D &= \frac{\sqrt2}{3}(\mathcal{M}_{10}^{11}-\mathcal{M}_{00}^{10})
	\end{split}
	\label{}
\end{equation}
Summing over the final and averaging over the initial spin states, the partial width is then given by

\begin{equation}
	\Gamma(A\to VP) = (|A_S|^2+|A_D|^2)PS_2
	\label{}
\end{equation}

For the $V$-decay into two pseudoscalar mesons $P_1$ and $P_2$ in the $P$-wave the decay amplitude will be given by

\begin{equation}
	\mathcal{M}_{00}^{1M}(\Omega_1)=N_1f_{00}^1D_{M,0}^{1*}(\Omega_1)
	\label{}
\end{equation}
where the helicity amplitude is $N_1f_{00}^1=\sqrt{3}a_P$.

Correspondingly, averaging over the $V$-spin states, the partial width is then given by

\begin{equation}
	\Gamma(V\to P_1P_2) = |A_P|^2 PS_2
	\label{}
\end{equation}

\section{Phase space convention}

The non-relativistic partial width is given by

\begin{equation}
	\Gamma(A\to BC)=2\pi|\mathcal{M}^{(NR)}_{A\to BC}|^2\times PS_2^{(NR)}
\end{equation}
where two-body non invariant phase space can be written as

\begin{equation}
	PS_2^{(NR)} = \int d^3\vec{k_B}d^3\vec{k}_C\delta^3(\vec{k}_B+\vec{k}_C)\delta(E_B+E_C-m_A)= 4\pi\frac{E_BE_Ck_C}{m_A}
\end{equation}

Since QPCM is in principle a non-relativistic model and we are using the relativistic Lorentz-invariant tensor formalism to describe $B\to K_1\gamma$ decay, one has to make some kind of continuation. In order to do that one has to

\begin{itemize}
	\item Use relativistic kinematics (i.e. $E_i^2=\vec{k_i}^2+m_i^2$).
	\item Use relativistic Breit-Wigner forms.
	\item Make the non-relativistic decay amplitudes to be ``relativistic'' correcting the phase space:
		\begin{equation}
		\Gamma(A\to BC) = \frac{1}{8\pi}\frac{k_C}{m_A^2}|\mathcal{M}^{(R)}_{A\to BC}|^2=8\pi^2\frac{E_BE_Ck_C}{m_A}|\mathcal{M}^{(NR)}_{A\to BC}|^2 \nonumber
		\end{equation}
		from where one immediately obtains the relation between the amplitudes
		\begin{equation}
			\mathcal{M}^{(R)}_{A\to BC}=8\pi^{3/2}\sqrt{E_BE_Cm_A}\mathcal{M}^{(NR)}_{A\to BC}
		\end{equation}
		Here $E_i$, $\vec{k}_i$ are the energies and momentum in the $A$-reference frame.
\end{itemize}

\vspace{1cm}

\newpage

\bibliographystyle{utphys}
\bibliography{bibliography}

\end{document}